\documentclass[preprint,showpacs,floatfix]{revtex4}
\usepackage{amssymb}
\usepackage{amsmath}
\usepackage[dvips]{graphicx}

\def\be{\begin{eqnarray} &&}
\def\nonu{\nonumber \\ &&}
\def\ee{\end{eqnarray}}

\def\bew{\begin{widetext}}
\def\ew{\end{widetext}}

\newcommand{\inv}[1]{{#1}^{-1}}

\newcommand{\mbf}[1]{\mathbf{#1}}

\newcommand{\lpro}{ ~_\ell \!\!\downarrow}
\newcommand{\rpro}{\downarrow _r~}
\def\Dslash{\raise.15ex\hbox{/}\kern-.7em D}
\def\Pslash{\raise.15ex\hbox{/}\kern-.7em P}

\usepackage{color}

\renewcommand{\bar}[1]{\overline{#1}}

\begin{document}
\title{Two-body scattering states in Minkowski space  and  the Nakanishi
integral representation onto the null plane}
\author{}
\author{T.   Frederico$^a$, G. Salm\`e$^b$ and   M. Viviani$^c$
 }
\affiliation{$^a$ Dep. de F\'\i sica, Instituto Tecnol\'ogico de Aeron\'autica,
DCTA, 12.228-900 S\~ao Jos\'e dos
Campos, S\~ao Paulo, Brazil\\
$^b$Istituto  Nazionale di Fisica Nucleare, Sezione di Roma, P.le A. Moro 2,
 I-00185 Roma, Italy \\
$^c$Istituto  Nazionale di Fisica Nucleare, Sezione di Pisa,
Largo Pontecorvo 3, 56100, Pisa, Italy }
\begin{abstract}
The Nakanishi perturbative integral representation of the
four-dimensional T-matrix is investigated in order to get
 a workable
treatment for  scattering states, solutions of the  inhomogeneous Bethe-Salpeter
Equation, in Minkowski space. The projection  onto the null-plane  of the four-dimensional
 inhomogeneous Bethe-Salpeter Equation plays a  key role for devising an  equation for
the Nakanishi weight
 function (a real function), as  in the homogeneous case that  corresponds
 to bound states and it has been already studied within  different frameworks.
   In this paper,
   the whole formal
  development is  illustrated in detail and applied to a
  system, composed by two massive scalars interacting through
  the exchange of
 a massive scalar. The explicit expression of the scattering integral equations  are also obtained
 in ladder approximation, and, as simple applications of our formalism,  some
 limiting cases, like  the zero-energy limit and
 the Wick-Cutkosky model in the continuum,  are  presented.

\end{abstract}%

 \maketitle



\section{Introduction}\label{intr}

The Bethe-Salpeter equation (BSE) \cite{SB_PR84_51} (see also
\cite{nakanishi} for an early and detailed review) is an
important tool for investigating
 the non perturbative regime of many important issues in physics, within a
relativistic framework. The huge tower of  difficulties for obtaining solutions has been bravely faced
with very different degrees of approximation and
with different mathematical tools, e.g. by exploiting  the classic Wick rotation trick \cite{WICK_54}
and free propagators (see e.g. Ref. \cite{Dork_2010}), by three-dimensional (3D) reductions
(see, e.g., Ref. \cite{Gross}) or by applying,  in Euclidean space, the combination of
 Dyson-Schwinger  and   Bethe-Salpeter equations (see e.g. Ref. \cite{DS} for
a recent review). The peculiar field of application  is  the description of
 bound states,  but one can find attempts for solving also the scattering-state
 problem (see, e.g., \cite{nishi,nakanishi,zemach,beyer}). As mentioned above, the number of approaches for
achieving solutions is quite large, and in this paper we focus on a particular
 scheme
 in
Minkowski space. In order to avoid the well-known difficulties related to the
cumbersome analytical behavior of the four-dimensional (4D) BSE, we
 adopt
the   framework proposed by
Nakanishi in his seminal work \cite{nak63}, the so-called perturbation theory
integral representation (PTIR) of i) the vertex function (that leads to the
BSE for bound states), ii) the  T-matrix (relevant for the present investigation
of the scattering states) and iii) the  general, multi-leg transition
amplitudes. He carefully studied the  parametric form of
any Feynman diagram that contributes to the multi-leg transition amplitudes for interacting
bosons, focusing on the denominator, since it governs the overall analytic behavior.

Nakanishi  was able, through a clever change of variable, to give a common form to any denominator appearing
in the parametric
expressions, and then he formally summed up the infinite series of Feynman 
diagrams.  The final result, {\it aside from the convergence problem of the perturbation
series} \cite{nak63},
makes it
possible to single out real, multi-variable weight functions (the Nakanishi amplitudes in what follows)
modulating
the denominator that generates
the analytic behavior in Minkowski space of  multi-leg transition amplitudes.
 It is noteworthy
   that a uniqueness theorem
  was proved for the PTIR transition
  amplitudes
  for bosonic  systems \cite{nak63}.

 Summarizing, PTIR yields a well-defined framework where
  classes of approximations
 to the multi-leg transition amplitude can be implemented in a systematic way,
 as shown, e.g., in  Refs. \cite{carbonell1,carbonell2}, where the BSE was
 investigated first within a ladder approximation of the kernel and then the
 cross-ladder class of diagrams was added.
 One could wonder if PTIR be able to deal with the analytic behavior of
 the non perturbative  multi-leg amplitudes. For the scattering case, 
 typical singularities
 like cuts, are certainly present in PTIR. For the bound states, one is
 interested in describing  the vertex functions, which is related to the residue of the T-matrix at
 the bound-state poles.  
   Moreover,  numerical investigations
 performed till now \cite{KW, carbonell1,carbonell2}
 have shown   that  PTIR is able to accurately obtain both  bound-state 
 BS amplitude and corresponding masses. This strongly indicates
 \cite{nak642} that
 solutions of the BSE, i. e.  eigenfunctions obtained in a non perturbative
 framework, can be put in the
 form suggested by Nakanishi PTIR.
   In nuclear and hadronic physics,
   the Nakanishi PTIR has been applied to
   a wide range of physical problems, but so far restricted to the bound-state case, 
   as shown, e.g., in Refs.
   \cite{Ji,KW,beyer,dae,sauli,carbonell1,carbonell2,carbonell3,carbonell4,
   carbonell5}.
The  relativistic scattering states play a relevant role in 
many hadronic processes, like in the light-meson decays of heavy hadron 
resonances (see, e.g., for a recent study \cite{patricia}), or even in  
more exotic cases (see e.g. \cite{giudice}). Therefore,  the extension of 
the Nakanishi PTIR for solving the BSE  in the continuum region,  
as one of the possible tools to deal with the 4D Minkowski structure
 without  approximations,
is appealing and deserves dedicated efforts.

Recently,  Carbonell and  Karmanov  proposed a new approach
for obtaining, within PTIR,  solutions of
 the BSE with truncated kernel, for bound states of both bosonic \cite{carbonell1,carbonell2,carbonell3} and fermionic
 \cite{carbonell4,carbonell5}
 two-body systems, interacting through the exchange of a massive scalar boson
 (Yukawa Model).
 As in the case of Ref. \cite{KW} (where only the bosonic case was investigated), they didn't make
 use of the Wick
rotation, but, for the first time, they  exploited   the covariant formulation of the
Light-front (LF) field theory \cite{cdkm}, obtaining a compact form for
the kernel of the BSE, in both ladder
\cite{carbonell1} and cross-ladder \cite{carbonell2} approximations.
Within such an approach, it is possible
to quickly get rid of the difficulties related to the analytic behavior of
the BS amplitude
in the Minkowski space, focusing the numerical efforts on determining a proper
Nakanishi weight function, that remarkably are  real functions, as
mentioned above.
The equation that determines such a weight function is directly related to
the 3D integral  equation for the valence component of the
interacting-system state. In turn, such a 3D equation is
 obtained by {\em projecting  onto the null-plane}  the full BSE, after expressing
the BS amplitude in terms of   the Nakanishi
PTIR. One may wonder how  the valence wave function be able to generate the whole
physical complexity of the full BS amplitude. Indeed,  within a non perturbative
framework, it formally turns out
\cite{sales00,sales01,hierareq,adnei07,adnei08,Tobias_FB} that by introducing
 the
projection onto the null-plane, namely the analytic integration over the variable $k^-=k^0-k^3$,
  the full
BS amplitude can be expressed in terms of  the valence component and   an
operator, producing the
richness  of the Fock states on top of  the lowest one. It must be emphasized
 that
such a  result has been  obtained for both bound and scattering
states of  bosonic and fermionic systems. This gives us confidence that the PTIR approach, adopted for
the bound state,  could  be extended to the scattering states.

Aim of this work is to construct the  integral equation that determines the
Nakanishi weight function for the   scattering states of two massive scalar bosons, exchanging
a massive scalar
 (cf \cite{carbonell1} for the bound state).
The main tool for obtaining the scattering equation is
the  projection onto the null-plane (LF-projection) of the BSE,
as developed in a series of papers
\cite{sales00,sales01,hierareq,adnei07,adnei08}. This method, not explicitly
covariant, can be applied
also to
the bound-state case, obviously  getting the same  result like in Ref.
 \cite{carbonell1}, where the explicitly-covariant LF framework is considered.
 Moreover, the new integral equation is applied to the
 simple ladder approximation for the Yukawa model, and  the extension  of 
  the Wick-Cutkosky model \cite{WICK_54,CUT_54} (a massless exchanged boson in
  ladder approximation)
  to the scattering region,  is proposed.

The  paper is organized as follows. In Sect. \ref{project}, it is introduced both
the general formalism of the BSE  and the integral transform adopted
for  projecting both  bound and scattering BSE's onto the
LF hyper-plane  (a short review on this topic can be found in Ref. \cite{Tobias_FB}). In Sect.
\ref{naka},
  a new  integral equation for determining the Nakanishi weight function
 for  4D scattering states is
  found, after briefly reviewing the bound state case. In Sect.
  \ref{seclad},  explicit expressions for both the  inhomogeneous term and
  the
  kernel in  ladder approximation are given for the  scalar Yukawa
  model with an exchanged massive  boson. Within such a framework,
  some relevant cases are discussed in Sect. \ref{applic}:
  i)  in subsect. \ref{zeroen},
  the zero energy scattering amplitude, ii) in subsect.  \ref{WiCu},
   the Wick-Cutkosky
  model for scattering states and  iii) in subsect. \ref{vbrevis}, a  new
  form for the integral equation determining the Nakanishi amplitude for bound
  states, in ladder approximation.
  Sect. \ref{concl} contains
concluding remarks. In
 Appendices  \ref{inho} and  \ref{vlad}
more details on the formal developments of Sect. \ref{seclad} are given.

\section{General formalism and LF projection of the BSE}
\label{project}

The ingredients of our approach are  in order: i) the PTIR of the BS amplitude,
 $\Phi(k,p)$, and ii) the LF projection of the BSE introduced for  obtaining
 a 3D equation for the valence wave function,
$\psi(\xi,\mbf{k}_{\perp})$, for both bound and scattering states. Once the 3D equation is
obtained, one can determine  the Nakanishi
amplitude, that in turn it yields the 4D BS amplitude within  the PTIR \cite{nak63}.
 It should be emphasized that
the LF projection of the BSE in Minkowski space represents a smart way to deal with
the singularities of the
4D $\Phi(k,p)$ once its analytic structure is known.

The {\em inhomogeneous} BSE for a scattering state of two scalar particles
 with equal mass $m$,  total
momentum $p=p_1+p_2$, total mass $M=\sqrt{p^2}$ and relative
momentum $k=(p_1-p_2)/2$  reads as follows
\be\label{bs}
\Phi^{(+)}(k,p,k_i)=(2\pi)^4\delta^{(4)}(k-k_i)+ G_0^{(12)}(k,p)
~
\int \frac{d^4k^\prime}{(2\pi)^4}i~{\cal K}(k,k^\prime,p)\Phi^{(+)}(k^\prime,p,k_i),
\ee
where $\Phi^{(+)}$ is the BS amplitude for the scattering state,
$i~{\cal K}$ the interaction kernel,
and $k_i$ the relative incoming momentum. It is worth noting that the matrix elements of the
T-matrix involved in
 the scattering process are the half-off-shell ones, since the incoming
 particles are on their mass shell.  This implies that the square of the
 intrinsic four-momentum is  given by
 $k^2_i= m^2 - M^2/4\le 0$ (cf Sect. \ref{naka}). The normalization of the free states is
\be
\langle k|q\rangle=(2\pi)^4\delta^{(4)}(k-q)
\ee
 and the free propagator of the two constituent particles is given by
\be\label{G0}
G_0^{(12)}(k,p) =G_0^{(1)}G_0^{(2)}
=\frac{i}{(\frac{p}{2}+ k)^2-m^2+i\epsilon}~~~
\frac{i}{(\frac{p}{2}-k)^2-m^2+i\epsilon} \ .
\ee

As well known, for a bound state one has the following  {\em homogeneous} integral equation
\be
\Phi_b(k,p)= G_0^{(12)}(k,p)
~
\int \frac{d^4k^\prime}{(2\pi)^4}i~{\cal K}(k,k^\prime,p)\Phi_b(k^\prime,p),
\label{bBSE}\ee
The above equations represent the starting point of our investigation.

In the rest of the Section, the integral transform adopted
for  projecting both  bound and scattering BSE's onto the
LF hyper-plane is described.
A reader  already acquainted of both the Bethe-Salpeter formalism and the LF
projection method can move  directly to Section~\ref{naka}.

\subsection{The BS amplitude and the LF wave-functions}
As  well known, the BS amplitude  is defined as the matrix
element between the vacuum $\langle 0|$ and a state $|p\rangle$, with total momentum $p^\mu$,
 of
the time ordered product of two Heisenberg operators. For a two-boson system, one has
\be\label{BSA}
\Phi(x^\mu_1,x^\mu_2,p^\mu)=\langle
0|T\left\{\varphi_H(x^\mu_1)\varphi_H(x^\mu_2)\right\}|p\rangle.
\ee
In general, the state vector $|p\rangle$ can be taken in different
representations, but  within  the LF quantization with only massive quanta,
it can be written in terms of  an infinite
  sum over
Fock components (see,  e.g. \cite{Brodsky:1997de}), i.e. the Fock vacuum
is an exact eigenstate
of the full LF Hamiltonian
(see also \cite{Brodsky010} for a recent discussion of the simplicity of the LF
vacuum). The weights of the Fock states in the infinite sum
are  the  LF wave functions (LFWFs), $\psi_{n/p}$, that have a probabilistic interpretation,
namely the proper integral over $|\psi_{n/p}|^2$ yields the probability of the corresponding
Fock state. It should be pointed out that the LFWFs provide an
intrinsic  representation of the  composite system, as a consequence of
the
possibility to separate the global motion from the intrinsic one, since the LF
boosts belong to the kinematical subgroup,  and the  mass operator is a
Lorentz scalar.

 In terms of the
four-momentum operator $\widehat p = (\widehat p^+, \widehat p^-,
\widehat{\mbf{p}}_{\!\perp})$, with $\widehat p^\pm = \widehat p^0 \pm \widehat p^3$,
the  invariant square mass operator
 reads (see e.g.\cite{Brodsky:1997de})
\be
{\widehat M}^2 = \widehat p^-\widehat p^+ - \widehat{\mbf{p}}^2_\perp \label{HLC}
\ee
The eigenvalue problem to be solved is
\be
{\widehat M}^2 \vert{p}\rangle = {\cal M}^2 \vert{p}\rangle,
\label{fullmassop}
\ee
where ${\cal M}^2$ is the  eigenvalue. The eigenstate, $\vert{p}\rangle$, can be
expanded  in multiparticle Fock
eigenstates $\{\vert{n} \rangle\}$ of the free LF
Hamiltonian, i.e  a basis constructed by using
the LF  plane waves, $\big\vert\tilde q \bigr\rangle$. The single particle free
state describes the
motion of a particle with mass $m$ and it is an eigenstate of the
LF momentum operators $\widehat p ^+$ and $\widehat{\mbf{p}}_{\!
\perp}$, i.e. $\tilde q \equiv \{ q^+,\mbf{q}_\perp\}$. The dependence upon the
minus component, given by $q^-=(m^2
+|\mbf{q}_\perp|^2)/q^+$, is understood in the notation we have adopted. The
free-state normalization is
\be
\bigl\langle \tilde k  \big\vert\tilde q \bigr\rangle = 2 k^+ (2 \pi)^3
\, \delta \bigl(k^ + - q^+ \bigr) \,\delta^{(2)} \negthinspace
\bigl(\mbf{k}_{ \perp} - \mbf{q}_\perp\bigr)= 2 k^+ (2 \pi)^3~\delta^3
\bigl(\tilde k - \tilde q\bigr)
\label{Pnorm}
\ee
with the completeness given by
\be
\int {d^3 \tilde q \over  2 q^+ (2 \pi)^3} ~\big\vert\tilde q \bigr\rangle~
\bigl\langle \tilde q  \big\vert
\ee
One should also recall that for a particle with mass $m$, one has
 \be
\int d^4q \delta(q^2-m^2)~\theta(q^0)= {1 \over 2} \int d^3 \tilde q \int dq^-~
\delta(q^+q^-
-|\mbf{q}_\perp|^2-m^2)~\theta(q^+) =
\nonu= \int {d^3 \tilde q \over 2 q^+} ~\theta(q^+)
\ee
where the positivity of $q^+$ follows from the positivity of $q^0$, since the
delta imposes that $q^0= [q^+ + (m^2+|\mbf{q}_\perp|^2)/q^+]/2$.

Actually,  each state $\vert p \rangle$  of the
composite system is expanded
in a complete
Fock basis of non interacting  $n$-particle states
$\vert n \rangle$  with $n\geq 2$, since we are dealing with the
BS amplitude of a composite two-boson system. Then one has
\be
\left\vert p \right\rangle = 2~(2\pi)^3~\sum_{n\ge 2}^\infty~\int \big[d \xi_i\big] \left[d^2
\mbf{k}_{i\perp }\right]\,
\psi_{n/p}(\xi_i,\mbf{k}_{i\perp })~ \bigl\vert n;~ \xi_i
p^+,~  \mbf{k}_{i\perp } + \xi_i \mbf{p}_{\! \perp} \! \bigr\rangle,
\label{LFstate}
\ee
where  the properly symmetrized free state
with $n$
particles are normalized, $\xi_i=q^+_i/p^+$ and
$\mbf{q}_{i \perp}= \mbf{k}_{i\perp }+ \xi_i ~\mbf{p}_{\perp }$, with
 $\mbf{k}_{i\perp }$ the intrinsic transverse momentum of the $i-th$
 constituent, $\psi_{n/p}(\xi_i,\mbf{k}_{i\perp })$ are the LFWFs (see below).
It should be pointed out that  $\{\xi_i,\mbf{k}_{i\perp }\}$
  is the set of intrinsic variables, fulfilling the
 following relations
 \be \sum_{i=1}^n \xi_i=1 \quad \quad \sum_{i=1}^n\mbf{k}_{i\perp} =0\ee
In Eq. (\ref{LFstate}),  the phase-space factors are given by
\be
\int \big[d \xi_i\big] \equiv \prod_{i=1}^n \int {d\xi_i\over 2 ~(2\pi) \xi_i } \,\delta
\Bigl(1 - \sum_{j=1}^n \xi_j\Bigr) ~,
\nonu
\int \left[d^2 \mbf{k}_{i\perp }\right] \equiv \prod_{i=1}^n \int
{d^2 \mbf{k}_{i\perp }\over (2\pi)^2}
\delta^{2} \Bigl(\sum_{j=1}^n\mbf{k}_{j\perp}\Bigr).
\label{phase}\ee
   Moreover, the kinematical nature of the LF
boosts
allows one  to write the overlap $\bigl\langle
n;~\xi_i p^+,~\mbf{k}_{i\perp } \xi_i \mbf{p}_{\perp }\big\vert p\bigr\rangle$
 in terms of the
global motion and  an intrinsic  function, i.e. the corresponding LFWF,
$\psi_{n/p}(\xi_i, \mbf{k}_{i\perp })$, viz
\be \label{eq:LFWF}
 \bigl\langle
n;~\xi_i p^+,~\mbf{k}_{i\perp }+ \xi_i \mbf{p}_{\perp }\big\vert p\bigr\rangle
= 2p^+ (2\pi)^3 \delta^3\Bigl(\sum_{i=1}^n \tilde q_i- \tilde p\Bigr)~\psi_{n/p}(\xi_i, \mbf{k}_{i\perp })=
\nonu=
2 (2\pi)^3 \delta \bigl(1 - \sum_{i=1}^n \xi_i \bigr) \,
\delta^{(2)}
\bigl(\sum_{i=1}^n \mbf{k}_{i\perp} \bigr)~\psi_{n/p}(\xi_i, \mbf{k}_{i\perp })
\ee
 For a bound system,  the  LFWFs are normalized according to
\be
\sum_n  \int \big[d \xi_i\big] \left[d^2 \mbf{k}_{i\perp }\right]
\,\left\vert \psi_{n/p}(\xi_i, \mbf{k}_{i\perp }) \right\vert^2 = 1.
\label{LFWFnorm}
\ee

  Thus, given the Fock
projection (\ref{eq:LFWF}), the state  of a composite system can be
determined  by  reaching through a suitable LF boost the intrinsic
frame, where $\mbf{p}_{\perp}=0$, and then solving the eigenvalue
problem for the invariant mass operator $\widehat M^2$, in order to
obtain the set of LFWFs.
 It should pointed out that if one applies a
dynamical LF rotation to the
 composite-system state, but still maintaining the Fock basis relative to the old
orientation of the hyperplane, then the Fock content of the state changes. The content
 returns to be
the same as before applying a dynamical rotation, if
one adopts the Fock basis corresponding to the new orientation of the hyperplane. This is
 analogous to the
Wigner transformations of the angular states, but generalized to the Fock space.

Summarizing, from Eqs. (\ref{BSA}) and  (\ref{LFstate}), one recognizes  that all the
Fock components with $n\ge 2$ are required for reconstructing
the BS amplitude in Minkowski-space. This is easily understood, since  the mass operator
(\ref{HLC}) is non-diagonal in the Fock space, due to the non commutativity of
the interaction  with respect to the particle-number operator.
However, only the valence
component
($n=2$) is active in the
calculation of the BS amplitude, (\ref{BSA}), if we restrict the analysis
 to
the  equal LF-times case ($x^+_1=x^+_2$) (see   below for  details).
\subsection{The BS amplitude and its Fock content}
The general picture deduced from Eqs. (\ref{BSA}) and  (\ref{LFstate}),
can be made more
complete, if we take into account the results obtained in Refs.
\cite{sales00,sales01,hierareq,adnei07,adnei08} for bound and scattering states of both
two-boson and
two-fermion systems.  There, by using the LF projection technique (see also \cite{Tobias_FB}),
it was shown
 that  the valence component of the LF bound and scattering states
contains enough information for reconstructing the  corresponding BS amplitudes.
It should be pointed out that  a similar   result for the bound state of both
two-boson and
two-fermion systems,
has been obtained    in
Refs. \cite{carbonell1,carbonell2,carbonell4}, but  by applying the Nakanishi PTIR
to the BS amplitude.

In what follows, the above mentioned relation between the valence component and the BS
amplitude is briefly illustrated.
From the field theoretical point of view, the states of the Fock basis,
 $\vert{n}\rangle$, are
 constructed
by applying free-field creation operators to the vacuum state
$\vert 0 \rangle$. Indeed, for a massive particle within the LF framework
(let us recall that the quantization
rules for the LF  field theory \cite{Brodsky:1997de} are given
at fixed LF time
 $\tau = t + z/c$ \cite{Dirac:1949cp}), the eigenvalue of $\widehat p^+ $ must be at least
 $m$, since the eigenvalues of $\widehat p^2$
 and $\widehat p^0$ are both
positive  (the mass of the particle yields the lower bound). Then,  the vacuum, that has
a vanishing eigenvalue of $\widehat p^+$, cannot  contain massive particles, given
 the conservation of the longitudinal momentum.
 This simple  kinematical chain makes the Fock
expansion  meaningful (for a new vistas on the condensate in QCD see Ref.
\cite{Brodsky010}). It should be pointed that the presence of the so-called
zero-modes, that can be associated to particles with vanishing masses, spoils the
triviality of the LF vacuum, i.e. $\widehat p^+ \vert 0
\rangle =0$ does not more entail a vanishing number of particles.
This means that the eigenvalue of $\widehat p^+$ is not sufficient for
identifying the vacuum with respect to a degenerate state populated by zero
modes. In this Section, we are considering a massive theory.

In the composite system we are considering, the
fundamental constituents appear in LF quantization as the
excitations of the dynamical fields,  $\varphi(x)$,
expanded in terms of bosonic creation and annihilation operators   on
the null-plane, with coordinates $x^- = x^0 - x^3$ and
$\mbf{x}_\perp$, at fixed LF time $x^+ = x^0 +
x^3$~\cite{Brodsky:1997de}.  For  $x^+=0$, the
Heisenberg operators turn into the Schr\"odinger
one, and therefore the field can be written in terms of non interacting
creation and annihilation operators as follows (notice that
$\varphi_H(x)= exp(iP^-x^+/2)~\varphi(\tilde x)~exp(-iP^-x^+/2)$)
\be\varphi(\tilde x)=  \int  \frac{d k^+}{\sqrt{ 2 k^+}}
\frac{d^2 \mbf{k}_\perp}{ (2 \pi)^{3/2}}~\theta(k^+ )~\left(a^{\dagger}(\tilde
k)e^{-i\tilde k\cdot \tilde x}+a(\tilde k)e^{i\tilde k\cdot \tilde
x}\right) \ ,
\ee
where $\tilde x=(x^-,\mbf{x}_\perp)$  and
$\tilde x \cdot \tilde k=k^+x^-/2 -\mbf{x}_\perp \cdot \mbf{k}_\perp$ (in this
notation $x^+=0$ is dropped off).
The  creation and annihilation operators satisfy the following commutation
relations
\be \label{eq:cr}
\left[a(\tilde k), a^\dagger(\tilde k')\right] =
 \,\delta (k^+ - {k'}^+)
\delta^{(2)}\negthinspace\left(\mbf{k}_\perp -
\mbf{k}'_\perp\right) .
\ee

A one-particle free state is defined by
$\vert \tilde k \rangle = (2 \pi)^{3/2}\sqrt{2 k^+}
\,a^\dagger(\tilde k) \vert 0 \rangle$, given the adopted normalization (cf Eq.
(\ref{Pnorm})), and $\langle \tilde x|\tilde k\rangle=
e^{i\tilde k\cdot \tilde x}$.

Let us  express $\Phi(x_1,x_2,p)$ in (\ref{BSA}) through its Fourier
transform. Translational invariance imposes to $\Phi$ the following
form
 \be
\Phi(x_1,x_2,p)=
\tilde{\Phi}(x,p) \;e^{-ip \cdot X},
\ee
where $X^\mu=(x^\mu_1+x^\mu_2)/2$,
$x^\mu=x^\mu_1-x^\mu_2$ and $\tilde{\Phi}(x,p)$ the reduced amplitude.
It is related to its  Fourier transform, $\Phi(k,p)$, as follows
 \be
\tilde\Phi(x,p)= \int {d^4k \over (2 \pi)^4}~e^{ik \cdot  x}~\Phi(k,p)
 \ee
where
\be p^\mu=p^\mu_1 + p^\mu_2 \quad \quad k^\mu={p^\mu_1-p^\mu_2 \over 2} \ee
with $p^2_i~\ne~m^2$ (see Eq. (\ref{bs})). The amplitude $\Phi(k,p)$ satisfies the homogeneous,  (\ref{bBSE}), or the inhomogeneous,
 (\ref{bs}),
BS equation depending if one is considering bound or
 scattering states.

For $X^\mu=0$, one has
\be\label{BSA1}
\tilde{\Phi}(x,p)=\langle
0|T\left\{\varphi_H(x^\mu/2)\varphi_H(-x^\mu/2)\right\}|p\rangle=
\nonu= \theta(x^+)~
\langle
0|\varphi(\tilde x/2)~e^{-i P^-  x^+/2}~ \varphi(-\tilde x/2)|p\rangle~
e^{ip^-  x^+/4}~+\nonu +~\theta(-x^+)~
\langle
0|\varphi( -\tilde x/2)~e^{i P^-  x^+/2}~ \varphi(\tilde x/2)|p\rangle~
e^{-ip^-  x^+/4 }=\nonu=
\theta(x^+)~
\sum_{n,n^\prime}e^{ip^-  x^+/4}\langle
0|\varphi(\tilde x/2)|n^\prime\rangle~
\langle n^\prime|e^{-i P^-  x^+/2} |n\rangle~\langle n|\varphi(-\tilde x/2)
|p\rangle~+ \nonu +~
\theta(-x^+)~
\sum_{n,n^\prime}e^{-ip^-  x^+/4}\langle
0|\varphi(-\tilde x/2)|n^\prime\rangle~ ~
\langle n^\prime|e^{i P^-  x^+/2} |n\rangle~\langle n|\varphi(\tilde x/2)|p\rangle
\label{BSAcov}\ee
where $|n\rangle$ and $|n^\prime\rangle$ are states of the Fock basis used in
 Eq. (\ref{LFstate}). It is easily realized that the matrix elements
 $\langle n^\prime|e^{-i P^-  x^+/2} |n\rangle$ are not diagonal in the Fock
 space, apart the case
 $x^+=0$, since the operator $P^-$ contains the interaction in its full glory.
Therefore,
 all the Fock components are acting in BS amplitude of the composite system.

\subsection{The BS amplitude and the valence wave-function}
The interesting case $x^+=0$ leads straightforwardly to the relation
between the valence component, $\psi_{n=2/p}$, and the BS amplitude
(see also \cite{cdkm}). In particular, one has
\be
\lim_{x^+\to 0_+}\tilde{\Phi}(x,p)=\langle
0|\varphi(\tilde x/2)~\varphi(-\tilde x/2)|p\rangle
\ee

If we perform the 4D Fourier transform of the quantity
$\tilde\Phi(x,p)~\delta(x^+)$ (i.e restricting the reduced amplitude to the LF
hypersurface, or equal LF-times case ($x^+= x^+_1-x^+_2=0$)), one can extract the valence wave function
$\psi_{n=2/p}(\xi, \mbf{k}_\perp)$. As a matter of fact, on one hand, one can write
\be
\int d^3 \tilde x~
e^{-i\tilde q \cdot \tilde  x}~\tilde\Phi(\tilde x,x^+=0 ,p)=
~{1 \over 2}\int {dk^- \over 2 \pi}\int {d^3 \tilde k\over (2 \pi)^3}~\Phi(k,p)
\int d^3 \tilde xe^{i(\tilde k -\tilde q)  \cdot \tilde x}=\nonu=
\int_{-\infty}^{\infty}{dk^-\over 2
\pi}~\Phi(k,p)~~~.
\label{LFpro}\ee
On the other hand, one has
\be
\int {d^3 \tilde x}~
e^{-i\tilde k \cdot \tilde  x}~\tilde\Phi(\tilde x,x^+=0_+ ,p)=
\int {d^3 \tilde x}~
e^{-i\tilde k \cdot \tilde  x}~\langle
0|\varphi(\tilde x/2)\varphi(-\tilde x/2)|p\rangle=\nonu=
\int {d^3 \tilde p_1 \over \sqrt{2 p^+_1}}
\int {d^3 \tilde p_2 \over \sqrt{2 p^+_2}}~2~\delta^3(\tilde k -{\tilde p_1
-\tilde p_2 \over 2})\int {d^3 \tilde q_1 \over 2 q^+_1 (2 \pi)^3}~
\int {d^3 \tilde q_2 \over 2 q^+_2 (2 \pi)^3}~2 p^+~ (2\pi)^3~\times \nonu
\delta^3(\tilde p -\tilde q_1
-\tilde q_2 )~
{1 \over \sqrt{2}}~\psi_{n=2/p}(\xi, \mbf{k}_\perp)~\sqrt{ 2 q^+_1}~\sqrt{ 2 q^+_2}~(2\pi)^3~
~\langle 0| a(\tilde p_1)~a(\tilde p_2)~a^\dagger(\tilde q_1)~
a^\dagger(\tilde q_2)|0\rangle=
\nonu ={ p^+\sqrt{2} \over ({p \over 2}+k)^+~({p \over 2}-k)^+}~
\psi_{n=2/p}(\xi, \mbf{k}_\perp)\ee
Summarizing, the valence component  is given by the Fourier transform
of $ \tilde\Phi(\tilde x, x^+=0_+,p)$, and it reads (cf \cite{cdkm})
\be
\psi_{n=2/p}(\xi, \mbf{k}_\perp)= {p^+\over \sqrt{2}}~\xi~(1 -\xi)~ \int_{-\infty}^{\infty}{dk^-\over 2
\pi}~\Phi(k,p)={p^+\over \sqrt{2}}~\xi~(1 -\xi)~\phi_{LF}(\xi, \mbf{k}_\perp)
\label{valwf}\ee
where $\xi=\xi_1$ and we have introduced the notation
$\phi_{LF}(\xi, \mbf{k}_\perp)$ for future purpose (see the next subsection).
 The expression (\ref{valwf}) can be considered as  a first bridge between
  the Fock state
decomposition of a composite-system state, Eq. (\ref{LFstate}), and
the BS amplitude. In particular, the integration over $k^-$ projects the BS amplitude
$\Phi(k,p)$ onto the LF-hyperplane, i.e. $x^+=0$, and it leads to the valence wave-function.

The LF
projection can be  directly applied  to both the  homogeneous and inhomogeneous  BSE (see
 Refs. \cite{carbonell1,carbonell2} for  the LF covariant approach for the bound state case).
 For instance, from Eq. (\ref{bBSE}) one has for the bound state
\be\label{BST} \int {dk^-\over 2 \pi}~\Phi_b(k,p)= \int {dk^- \over
2 \pi}~G_0^{(12)}(k,p) \int\frac{d^4k^\prime}{(2\pi)^4} i~{\cal
K}(k,k^\prime,p)\Phi_b(k^\prime,p), \ee This mathematical step
is a fundamental one in order to get  an equivalent
equation that allows one to determine the
Nakanishi amplitude, eventually leading
 to the BS equation treated  within PTIR framework\cite{nak63},
(as discussed in detail in the next
Section). In what follows, the LF projection is applied to both bound and scattering valence
wave-functions.

\subsection{The LF projection of the  BS equation}

The LF projection,  applied to the BS amplitudes for bosonic and fermionic states,
 can be embedded in a more general
framework \cite{sales00,sales01,hierareq,adnei07,adnei08}, where the
relation (\ref{valwf}) can be  extended to the scattering BSE without relying on
 any
perturbative approach: we reiterate that
 this represents the main motivation of the present paper. It should be also mentioned that
 the general formalism of Refs.\cite{sales00,sales01,hierareq,adnei07,adnei08} allows one to
  fully reconstruct the 4D BS
amplitude  in Minkowski space from the 3D valence component  (cf the following Eqs. (\ref{invop})
and (\ref{psitolfa})).

Let us briefly review the results for  an interacting two-boson system \cite{sales00,hierareq,adnei07}
(see  Refs. \cite{sales01,adnei08} for the
generalization to a fermionic system and  Ref. \cite{Tobias_FB} for an introduction to the three-boson
 case). In order to get the 3D eigenfunction for the valence component from the 4D
 BSE,
   the LF projection technique  is combined with a Quasi-potential
 treatment of the BS interacting kernel (see \cite{wolja}). This new ingredient
makes it
 possible  to single
 out from the 4D  Green's function of the interacting system the "trivial"
 global propagation. Such a formal step is accomplished by using the LF projection, i.e.
 the proper integration over the $k^-$ variable. Then, one remains with the valence dynamics
 described by a  3D effective  LF interaction directly related to the BS kernel, through
 suitable projection operators.

 The main outcome for  bound and scattering states is given by the
 3D LF eigenequation for the valence component that can be obtained by applying the LF-projection
 technique
  to the BS equation. This can be rewritten as follows
 \be
 G^{-1}(p)~|\Psi\rangle= \left[G^{-1}_0(p)-V(p)\right]~|\Psi\rangle=0
 \label{bseq}\ee
  where an overall delta function for the total four-momentum conservation has been factorized out,
  $G_0(p)$  is the free
  propagator and $V(p)$ is the 4D interaction,
  obtained from some given  Lagrangian  \cite{zuber}.Their  matrix elements are given by
  \be
  \langle p_1|G_0(p)| p'_1\rangle=\delta^4(p^{\prime}_1-p_1)
~G_0^{(12)}(k,p)
\nonu
\langle p_1|V(p)| p'_1\rangle=~i{\cal K}(k,k',p)~~~~~~.
  \ee
  where $p=p_1+p_2=p'_1+p'_2$,  $k=(p_1-p_2)/2$ and $k'=(p'_1-p'_2)/2~$.
   Then, following
 Refs. \cite{sales00,sales01,hierareq,adnei07,adnei08} and the resum\'e
 in Ref. \cite{Tobias_FB}, one obtains the following 3D LF eigenequation (that holds
 for bound and scattering states, depending upon the boundary conditions)
\be
 ~\langle k^+,\mbf{k}_\perp|g^{-1}_0(p )-w(p)~| \psi_{n=2/p}\rangle=0
\label{valeneq}
\ee
where
the 3D free Green's function,  $g_0(p)$ describes the global propagation
and  $w(p)$ is the effective 3D interaction, related to the 4D
Quasi-potential (cf \cite{wolja}).
\be W(p)=V(p)+V(p)\Delta_0(p)W(p)\ee
where the key ingredients is contained in
$\Delta_0(p)=G_0(p)-\tilde G_0(p)$ and it is represented by  the auxiliary Green's function
$\tilde G_0(p)$, that is nothing else  but the 4D image of the 3D $g_0(p)$. Through
$\Delta_0(p)$, one can factorize the intrinsic free motion, after subtracting the
"trivial" global one.  The 3D free Green's function is defined by\be
g_0(p)=\lpro G_0(p) \rpro \ee
where  the two new symbols $  \lpro $ and $\rpro$, that indicate
 a projection onto the LF hyperplane $x^+=0$,
  are given by
\be \lpro  ~\equiv~ \int {dk^{ -}_1 \over 2\pi}~\langle k^{
-}_1|~~, \quad \quad \quad  \rpro ~ \equiv~ \int {dk^{ -}_1 \over 2\pi}~
| k^{ -}_1 \rangle \ee
Then, 3D quantities in Eq. (\ref{valeneq}) have the following expressions 
\be\inv{g_0(p)}=\overline\Pi_0(p)~\left[G_0(p)\right]^{-1}~\Pi_0(p)
 \nonu w(p):= {\overline \Pi}_{0}(p) ~W(p)~ \Pi_{0}(p) \label{3Df} \ee
where the {\em free reverse LF projection operator}  and its LF-conjugated, are
\be \Pi_0(p):= G_0(p)  \rpro ~\inv{g_0(p)}~~, \quad \quad\quad \quad
\overline\Pi_0(p):=\inv{g_0(p)}~ \lpro   G_0(p).
\label{freeprojt}\ee
 Finally, the 4D auxiliary Green's function
reads \be \tilde G_0(p)= \Pi_{0}(p)~g_0(p)~\overline\Pi_0(p) \ee It
has to point out that $\Pi_0(p)$ action is $3D\to 4D$, while
$\overline\Pi_0(p)$ acts in the reverse way.
The explicit expression of $g^{-1}_0(p )$ is
\be g^{-1}_0(p) ~= ~-i
~\theta(\widehat k_1^+)~\theta(p^+ -\widehat k_1^+) ~\widehat
k_1^+~(p^+-\widehat k_1^+)~ \left[p^- -\widehat k^-_{1on} - \widehat
k^-_{2on}+i\epsilon\right]~. \label{g0m2}
\ee
where
\be
\widehat k^-_{1on} |\tilde k_{1}\rangle=
{m^2 +|\mbf{k}_{1\perp}|^2\over k_{1}^+}~
 |\tilde k_1\rangle
\ee
and the analogous for particle $2$.
The  diagrammatic content (in the Fock space) of the Quasi-potential can
be argued from the following formal solution of the integral equation that
determines $W(p)$
\be
W(p)=V(p)+V(p)\Delta_0(p)W(p)=V(p)\sum_{N=0}^\infty\left[\Delta_0(p)V(p)\right]^N\
, \label{2.3a}
\ee
where the $\Delta_0(p)$ insertions  bring information
 on the
intermediate particle propagations. A corresponding analysis can be
performed for the LF effective interaction $w(p)$, and it is
physically quite transparent, since it can be  carried out within a
LF-time ordered framework. In particular, $w(p)$ contains all
possible LF-time ordered exchanges, corresponding to a sum over
diagrams with any number of intermediate particles: this happens
even for the ladder BSE. It should be pointed out that the ladder
approximation of the LF eigenequation, Eq. (\ref{valeneq}), where
$w(p)$ is approximated by the first term in the power expansion in
Eq. (\ref{2.3a}), does not account for the full complexity of the
 4D ladder BSE. The physical reason lies in the fact that the
iterations of the  4D ladder kernel  and the 3D LF ladder
kernel (time-ordered graphs) generate different intermediate states.
The LF ladder kernel and its iterations contain
 only one exchanged particle (at a given global LF-time ) in
the intermediate state, whereas the
iterations of the ladder 4D kernel contain also many-body
states, with increasing number of exchanged particles (stretched
boxes, see e.g.  \cite{bakker}). This leads to a difference in the binding energies, which
is however small \cite{sales00,miller,mariane}.  In principle, for any
4D kernel given by a finite set of irreducible
graphs, both BS (\ref{bseq}) and LF   equation, Eq. (\ref{valeneq}), gives the same
eigenvalue, once $w(p)$ comes from the solution of (\ref{2.3a}).

In the Quasi-potential framework the {\em interacting LF
 reverse projection operator} (see Ref. \cite{adnei08})
 \be \Pi(p)=~\left [1 +\Delta_0(p)~W(p)~\right ]~ \Pi_{0}(p)\label{invop}\ee
leads to the following relations between the 3D valence component
and the 4D BS amplitude
 \be~ | \Psi  \rangle=~\Pi(p)~ | \phi_{LF}
\rangle\label{psitolfa} \\ && \lpro ~ |\Psi  \rangle=~
|\phi_{LF} \rangle\label{psitolfb}.\ee
The 3D valence wave-function, $\phi_{LF}$, has been already introduced
in Eq. (\ref{valwf}).
The analogous relations for  the free case read
 \be~ | \Psi_0  \rangle=~\Pi_0(p)~ | \psi_{0}
\rangle\label{psitolfr} \\ && \lpro ~ |\Psi_0  \rangle=~
|\psi_{0} \rangle\label{psitolfs}.\ee

It turns out that the full
complexity of the Fock-space in the 4D BS amplitude appears, not
only through the effective interaction that determines the valence
wave function, but also through the interacting reverse projection
operator, $\Pi(p)$.

Finally, let us remind the 3D integral equations that follow from
 the eigenequation (\ref{valeneq}), for both the bound and
the scattering states, viz
\be
\phi^b_{LF}(\xi, \mbf{k}_\perp) =
 ~\langle k^+,\mbf{k}_\perp|g_0(p )w(p)~| \phi^b_{LF}
\rangle~~~~,
\nonu
\phi^{(+)}_{LF}(\xi, \mbf{k}_\perp) =
 ~\langle k^+,\mbf{k}_\perp|1+g(p )w(p)~| \psi_0
\rangle~~~~,
 \label{solval}
\ee
with $g^{-1}_0(p ) ~|\psi_0\rangle=0$.

\section{Integral equation for the Nakanishi weight function for bound and
scattering states}
\label{naka}
In this Section, we illustrate how to obtain an integral
equation for the
Nakanishi weight functions for scattering states, starting from a BS equation
for a system composed by two massive
scalars,  exchanging  a  scalar particle. The kernel is composed
by the infinite
sum of two-particle
irreducible diagrams \cite{zuber,grossbook}, and the self-energy contribution to
the massive two-particle propagation are discarded, at the present stage.

In order to have a
suitable introduction to the scattering case, we first   briefly discuss
 the S-wave bound-state case, within our LF approach. The same integral equation
 for the Nakanishi weight function has been devised by
  Carbonell and Karmanov
\cite{carbonell1,carbonell2}, but
within the explicitly covariant LF approach \cite{cdkm}.

\subsection{The bound states} In this subsection the S-wave bound-state integral
equation for determining the Nakanishi weight function is presented by using
 the
LF language introduced in the previous Section (namely a non explicitly covariant
formalism). The BS amplitude is
written
in terms of the Nakanishi PTIR
\cite{nakanishi,nak63,carbonell1}, $g_b(\gamma',z';\kappa^2)$, as follows
\be\label{bsint}
\Phi_b(k,p)=~-i~\int_{-1}^1dz'\int_0^{\infty}d\gamma'
~
\frac{g_b(\gamma',z';\kappa^2)}{\left[\gamma'+m^2
-\frac{1}{4}p^2-k^2-p\cdot k\; z'-i\epsilon\right]^{2+n}}=
\nonu =
~i (-1)^n~\int_{-1}^1dz'\int_0^{\infty}d\gamma'
~
{g_b(\gamma',z';\kappa^2)\over\left[{k}^2+p\cdot k
z'-\gamma'-\kappa^2+i\epsilon\right]^{2+n}}
\ee
where $p^2=M^2$ is the invariant mass of the interacting system and
 \be\label{kappa2}
\kappa^2 = m^2- {M^2\over 4}.
\ee
with $m$ the constituent mass.
For bound states $\kappa^2 > 0$, while for the scattering ones $\kappa^2\le
0$.
In  general, the power $n$ in  the denominator
is a dummy integer
parameter, and  in Refs. \cite{carbonell1,carbonell2} the  minimal value $n=1$ was
chosen for the sake of simplicity.
Bigger values of $n$ may result in a more smooth solution for
$g_b(\gamma',z';\kappa^2)$
\cite{KW,carbonell5}.

 The LF projection allows one to readily get rid of the singularity in Eq.
 (\ref{bsint}), eventually
 obtaining the valence wave function, within the PTIR framework.
 In agreement with
Ref. \cite{carbonell1}, one can define $\psi_b = \sqrt{2}~ \psi_{n=2/p}$
as follows (cf Eq. (\ref{valwf}))
\be \label{lfwf3a}
\psi_b(\xi,k_\perp)=~p^+~\xi~(1-\xi)
\int {dk^- \over 2 \pi}
\Phi_b(k,p)=\nonu
=~\xi~(1-\xi)~\int_0^{\infty}d\gamma'~\frac{
g_b(\gamma',1-2\xi;\kappa^2)}
{[\gamma'+k_\perp^2 +\kappa^2+\left(2\xi-1\right)^2 {M^2\over 4}]^2}.
\ee
It should be pointed out that the denominator is always positive, and one can
safely drop out the $i\epsilon$ term.

In order to strictly follow the notation of Ref. \cite{carbonell1},  let us use
 the  variables ($\gamma,z$), related to the standard LF momentum
as follows:
\be \gamma=k_{\perp}^2 \quad \quad   1\geq z=1-2\xi\geq -1 ~~~~.\ee
 Then,  one can rewrite
\be \label{lfwf3a1}
\psi_b(z,\gamma)
={(1-z^2)\over 4}~\int_0^{\infty}d\gamma'~\frac{
g_b(\gamma',z;\kappa^2)}
{[\gamma'+\gamma +z^2m^2+(1-z^2)\kappa^2]^2}
\ee
The integral equation for the Nakanishi weight function,
$g_b(\gamma, z;\kappa^2)$, is obtained by inserting (\ref{bsint}) in both
sides of (\ref{BST}),  and using Eqs. (\ref{lfwf3a}) and  (\ref{lfwf3a1})(see
Appendix A of \cite{carbonell1}) for  details). One gets
\be
\int_0^{\infty}d\gamma'~\frac{g_b(\gamma',z;\kappa^2)}{[\gamma'+\gamma
+z^2 m^2+(1-z^2)\kappa^2]^2} =
\int_0^{\infty}d\gamma'\int_{-1}^{1}dz'\;V^{LF}_b(\gamma,z;\gamma',z')
g_b(\gamma',z';\kappa^2).
\label{ptireq}\ee
where the kernel $V_b$, appearing in the right-hand side of Eq.
(\ref{ptireq}), is
related to the kernel $i{\cal K}$ in the BS equation, (\ref{bs}), as follows
\be\label{V}
V^{LF}_b(\gamma,z;\gamma',z')=~i
p^+~\int_{-\infty}^{\infty}{d k^- \over 2\pi}~G_0^{(12)}(k,p)
\int \frac{d^4k'}{(2\pi)^4}\frac{i{\cal K}(k,k',p)}
{\left[{k'}^2+p\cdot k' z'-\gamma'-\kappa^2+i\epsilon\right]^3}
=\nonu=
~-ip^+\int_{-\infty}^{\infty}{d k^- \over 2\pi}~
{1 \over \left[(\frac{p}{2}+k)^2-m^2+i\epsilon\right]}
~ {1\over \left[(\frac{p}{2}-k)^2-m^2+i\epsilon\right]} \times
\nonu
\int \frac{d^4k'}{(2\pi)^4}\frac{i{\cal K}(k,k',p)}
{\left[{k'}^2+p\cdot k' z'-\gamma'-\kappa^2+i\epsilon\right]^3}
\label{vbou}\ee
Its explicit expression within both  the ladder and the cross-ladder
approximation
for the kernel ${\cal K}$ can be found in Refs. \cite{carbonell1,carbonell2}.

We emphasize that Eq. (\ref{ptireq}) is
equivalent to the initial BS equation (\ref{bs}), within the PTIR framework.
Moreover, it should be pointed out that the  mass,
$M$, of the interacting system appears on both sides of Eq. (\ref{ptireq})
and is contained in the parameter $\kappa^2$. Once the weight function
$g_b(\gamma',z;\kappa^2)$ is obtained, then we can reconstruct the BS amplitude
 (see Eq. (\ref{bsint})). As already mentioned in the
 Introduction, the ability of the valence wave function to reconstruct the
 full BS amplitude is not only restricted to a perturbative framework, but
 it holds also in a non perturbative analysis, and even more for both bound and
 scattering states
 \cite{sales00,sales01,hierareq,adnei07,adnei08}. This observation
 represented our
 motivation to face with the investigation of the scattering sates.

 Finally some important remarks. It turns out  \cite{KW,carbonell1} that
 $g_b(\gamma,z;\kappa^2)$ may be
zero in an interval $0\le \gamma \le \gamma_0$, and that  the exact value
where it differs from zero is determined by Eq.
(\ref{ptireq}) itself. Moreover,  the uniqueness of the Nakanishi
weight function is ensured
by a theorem (see \cite{nak63}). Following Refs. \cite{nak642,KW}, one
can extract from   (\ref{ptireq}) the following eigenequation
\be
g_b(\gamma,z;\kappa^2)=~\int_{0}^{\infty}d\gamma'\int_{-1}^{1}dz'\;
{\cal V}_b(\gamma,z;\gamma',z';\kappa^2)
g_b(\gamma',z';\kappa^2)
\ee
 For a discussion of the above equation within the ladder-approximation
framework, see
subsect. \ref{vbrevis}.

\subsection{The scattering states}
The LF-projection technique, combined with the PTIR
framework, has the appealing feature of
 systematically removing  singularities, that appear in  both   BS
 amplitude and
 kernel, by
integrating
  over $k^-$  more simple analytical structures.  We notice that
  the quadratic $k^0$ dependence in the propagators
  translates into a linear dependence upon $k^-$  (see, e.g.
  \cite{bakker1,Saw}). Since  within  PTIR approach,  the analytical
 form  of any
multi-leg amplitude is made explicit, the linearity of the pole structure
attained through the LF technique is advantageous.
Then, it is natural to extend the
investigation to the
scattering case, increasing the complexity of the problem.

 According to the general treatment of  a four-leg amplitude, as   developed in Ref.
\cite{nakanishi,nak63},  one   can write
 the off-shell T-matrix as follows
\be
\langle k^{\prime \mu}|T(p)|k^{ \mu}\rangle= \Pi_{i=1,7} \int_0^1
dz_i~\delta \left (1-\sum_{j=1,7}z_j\right )~\int_0^\infty d\gamma ~\times
\nonu {{\cal G}^+(\gamma,z_i)
\over \gamma- \left[z_1\left( p/2+ k^\prime\right)^2 +z_2\left (p/2-k^\prime\right)^2
+z_3\left( p/2+k\right)^2 +z_4\left (p/2-k\right)^2+z_5 ~s +z_6
t+z_7u\right]-i\epsilon}\nonu\label{fulloff}\ee
where the Mandelstam variables are
\be
s=p^2 \quad \quad t=(k^\prime-k)^2 \quad \quad u=(k^\prime+k)^2\ee
and they satisfied the four-momentum conservation, that reads
\be
s+t+u=p^2+2(k^2 +k^{\prime 2})
\label{Mandel}\ee
Equation (\ref{fulloff}) has a redundant dependence upon the seven invariant squares, 
in
order to have a compact form of the fully-off-shell four-leg amplitude (namely,
without a decomposition in $s$, $t$ and $u$ channels). It is
understood that the actual dependence is upon a set of six independent 
variables, given the constraint in Eq. (\ref{Mandel}).

Indeed, for evaluating the scattering states we need half-off-shell T-matrix,
i.e. one has to put on their-own mass shell the two incoming particles with
four-momenta $p/2 \pm k_i$. Then one has
\be
\langle k^{ \mu}|T(p)|k^{ \mu}_i\rangle= \Pi_{i=1,7} \int_0^1
dz_i~\delta \left (1-\sum_{j=1,7}z_j\right )~\int_0^\infty d\gamma ~\times
\nonu {{\cal G}^+(\gamma,z_i)
\over \gamma- \left[z_1\left( p/2+ k\right)^2 +z_2\left (p/2-k\right)^2
+z_3m^2 +z_4m^2+z_5 ~p^2 +z_6
t+z_7u\right]-i\epsilon}
\label{half1}\ee
where $( p/2+k_i)^2=( p/2-k_i)^2=m^2$. As a consequence $p\cdot k_i=0$, and
therefore
$k^2_i=m^2-p^2/4=\kappa^2\leq0$.

Let us analyze in detail the $z$-dependent part of the
denominator in Eq. (\ref{half1}). By exploiting
the presence of the delta function in Eq. (\ref{half1}) for eliminating
$z_5$, it can be rewritten as follows\be
z_1\left( p/2+ k\right)^2 +z_2\left (p/2-k\right)^2
+(z_3 +z_4)m^2+z_5 ~p^2 +z_6
t+z_7u=
\nonu=(2-2z_1-2z_2-2z_3-2z_4  -3z_6-3z_7){p^2\over 2}+
(z_1+z_2+z_3+z_4+2z_6+2z_7)m^2
+\nonu+(z_1+z_2+z_6+z_7)~\left[(k^2+{p^2\over 4}-m^2)
+\zeta^\prime p\cdot k
+  \zeta 2 k\cdot k_i\right]
\ee
where $\zeta=(z_7-z_6)/(z_1+z_2+z_6+z_7)$ and
$\zeta^\prime=(z_1-z_2)/(z_1+z_2+z_6+z_7)$ belong to $[-1,1]$. Remind that
the variables $z_i$ fulfill i)
 $\sum_{i=1,7}z_i=1$ and ii) $0\le z_i\le 1$. Furthermore
by introducing a new variable
$$\gamma''={\gamma +\left({p^2}-m^2\right)(z_1+z_2+z_3+z_4+2z_6+2z_7)
\over(z_1+z_2+z_6+z_7) }
-{p^2\over 2}{(2-  z_6-z_7)\over(z_1+z_2+z_6+z_7) }$$ that belongs to
$[-\infty,\infty]$, recalling that  $(z_1+z_2+z_6+z_7)\in[0,1]$, then one
  can rewrite the half off-shell
T-matrix as follows \be \langle k^{ \mu}|T(p)|k^{ \mu}_i\rangle=
\int_{-1}^1 d\zeta\int_{-1}^1 d\zeta^\prime \int_{-\infty}^\infty
d\gamma'' ~ {\widetilde{\cal
G}^+(\gamma'',\zeta,\zeta^\prime) \over
\gamma''-\left[k^2+{p^2\over 4}-m^2 +\zeta^\prime p\cdot k +
\zeta 2 k\cdot k_i\right] -i\epsilon} \label{half2} \ee where the
weight function, $\widetilde{\cal
G}^+(\gamma'',\zeta,\zeta^\prime)$, is the result of multifold
integrations and changes of variables, as suggested by the previous
calculation.
The scattering wave can be obtained by considering the
following vertex function \be \langle k^{ \mu}|G_0(p)T(p)|k^{
\mu}_i\rangle= {i \over \left ({p \over 2} +k\right )^2-m^2+i
\epsilon}~{i \over \left ({p \over 2} -k\right )^2-m^2+i \epsilon}
\times \nonu \int_{-1}^1 d\zeta\int_{-1}^1 d\zeta^\prime
\int_{-\infty}^\infty d\gamma'' ~ {\widetilde{\cal
G}^+(\gamma'',\zeta,\zeta^\prime) \over \left[k^2+{p^2\over
4}-m^2 +\zeta^\prime p\cdot k +  \zeta 2 k\cdot k_i\right]
-\gamma'' +i\epsilon}= \nonu =~\int_{-1}^1 d\zeta\int_{-1}^1
d\zeta^\prime \int_{-\infty}^\infty d\gamma'' ~ \int_0^1
d\alpha_1 \int_0^{1-\alpha_1}  d\alpha_2~ {\widetilde{\cal
G}^+(\gamma'',\zeta,\zeta^\prime) \over \left[D
+i\epsilon\right]^3} \ee
where the Feynman trick
\be
{1 \over ABC}=\int_0^1 d\alpha_1 \int_0^{1-\alpha_1}  d\alpha_2~{1 \over
\left[\alpha_1 A+
\alpha_2 B +(1-\alpha_1-\alpha_2)C +i\epsilon\right]^3}
\ee
has been used.
Hence,  the denominator $D$ is given by \be
D=(1-\alpha_1-\alpha_2)\left[k^2+{p^2\over 4}-m^2 +\zeta^\prime
p\cdot k +  \zeta 2 k\cdot k_i -\gamma''\right] +\nonu
+\alpha_1
 \left[ \left ({p \over 2} +k\right )^2-m^2\right]+\alpha_2
 \left[ \left ({p \over 2} -k\right )^2-m^2\right]=
 \nonu=
 k^2+{p^2\over 4}-m^2 +p\cdot k~z''
 +2 k\cdot k_i~z' -\gamma^\prime
 \ee
 with $z''=\zeta^\prime
 (1-\alpha_1-\alpha_2) + \alpha_1-\alpha_2$ and   $z'=\zeta
 (1-\alpha_1-\alpha_2)$ belonging to $[-1,1]$ and
 $\gamma^\prime=\gamma''(1-\alpha_1-\alpha_2)$.
 Then,  the vertex function becomes
 \be
\langle k^{ \mu}|G_0(p)T(p)|k^{ \mu}_i\rangle=~-i~
\int_{-1}^1 dz' \int_{-1}^1
dz ''
\int_{-\infty}^\infty d\gamma' ~
{g^{(+)}(\gamma',z',z'')
\over \left[k^2+{p^2\over 4}-m^2 +p\cdot k~z''
 +2 k\cdot k_i~z' -\gamma' +i\epsilon\right]^3}\nonu
\label{g0t}\ee
where a factor $(-i)$ has been inserted for convenience (cf \cite{carbonell1}
for the bound state) and
\be
g^{(+)}(\gamma',z',z'')=~i~\int_0^1 d\alpha_1 \int_0^{1-\alpha_1}  da
~{1 \over a^3} ~\widetilde{\cal
G}^+\left({\gamma'\over a} ,{z'\over a},
{z''\over a}\right)\ee
with $a= 1-\alpha_1-\alpha_2$. The vanishing behavior of $\widetilde{\cal
G}^+$ for  values of its arguments at infinity makes finite the twofold integration.

A similar representation can be introduced for the scattering
state by using Eq. (\ref{g0t}), with the same power as in the case of the bound state analyzed in Refs.
  \cite{carbonell1,carbonell2}, viz \be \label{bsintscat}
\Phi^{(+)}(k,p)= (2\pi)^4\delta^{(4)}(k-k_i)+ \nonu ~-
i~\int_{-1}^1dz'\int_{-1}^1dz''\int_{-\infty}^{\infty}d\gamma'
 \frac{g^{(+)}(\gamma',z',z'')}{\left[\gamma'+m^2
-\frac{1}{4}M^2-k^2-p\cdot k\; z''-2 k\cdot k_i~z' -i\epsilon\right]^3}=\nonu=
(2\pi)^4\delta^{(4)}(k-k_i)-
i~\int_{-1}^1dz'\int_{-1}^1dz''\int_{-\infty}^{\infty}
d\gamma'
 ~\times \nonu\frac{g^{(+)}(\gamma',z',z'')}{\left[\gamma'+\gamma+\kappa^2
-k^-(k^+ +{M \over 2}z''-{M \over 2}z_i z') -k^+( {M \over 2}z'' + k^-_i z') +
 2 z'cos \theta \sqrt{\gamma
\gamma_i}-i\epsilon\right]^3}
\label{ptirsc}\ee
where we adopt the frame $\tilde p=\{M,{\bf 0}\}$. Following
the formalism of Ref. \cite{carbonell1,carbonell2},
$z_i=-2 k^+_i/M$ ($1\geq |z_i|$ since the incoming particles have positive
longitudinal momenta, i.e. $p^+/2 \pm k^+_i \geq 0$), $cos \theta =\widehat {\bf k}_\perp \cdot
\widehat{\bf k}_{i\perp}$,
$\gamma=|{\bf k}_\perp|^2$ and $\gamma_i=|{\bf
k}_{i\perp}|^2$.
 It should be pointed out that for the variable $\gamma'$ a lower bound
different from $-\infty$  could be possible, as discussed in Sect. \ref{seclad}
 (below Eq. (\ref{gladder})).

The 3D
LF scattering wave function is given by ($\psi^{(+)}=\sqrt{2}~
 \psi^{(+)}_{n=2/p}$) \be
\psi^{(+)}\left(z,\gamma,cos\theta\right)= p^+ {(1-z^2)\over
4} \int {dk^-\over 2 \pi}~ \Phi^{(+)}(k,p)=\nonu= p^+ {(1-z^2)\over
4} ~(2\pi)^3\delta^{(3)}(\tilde k-\tilde k_i) + {(1-z^2)\over  4}
\int_{-1}^1 dz'~\times \nonu
\int_{-\infty}^{\infty}d\gamma'\frac{g^{(+)}(\gamma',z',z;\gamma_i,z_i)}
{[\gamma'+\gamma+ z^{ 2} m^2+(1- z^{ 2})\kappa^2
 +{M \over 2} z~z' ({M \over 2} z_i + k^-_i ) +
 2 z'cos \theta \sqrt{\gamma
\gamma_i}-i\epsilon]^2}\label{scat1}\ee
where the dependence upon the initial
variables $\{\gamma_i,z_i\}$ in $g^{(+)}$ has been made explicit, for the sake of
clarity.
The result in Eq. (\ref{scat1}) can be obtained by integrating  over $k^-$ the
singular integral in Eq. (\ref{ptirsc}) as follows
(cf Appendix A in Ref. \cite{carbonell1})
\be
\int_{-\infty}^\infty {dk^-\over 2 \pi}{1 \over \left[\gamma'+\gamma+\kappa^2
-k^-(k^+ +{M \over 2}z''-{M \over 2}z_i z') -k^+( {M \over 2}z'' + k^-_i z') +
 2 z'cos \theta \sqrt{\gamma
\gamma_i}-i\epsilon\right]^3}=\nonu=
{i \over 2}{\delta(k^+ +{M \over 2}z''-{M \over 2}z_i z') \over
 [\gamma'+\gamma+\kappa^2
 -k^+( {M \over 2}z'' + k^-_i z') +
 2 z'cos \theta \sqrt{\gamma
\gamma_i}-i\epsilon]^2}=\nonu
={i \over M}{\delta(- z  +z''-z_i z') \over [\gamma'+\gamma+ z^{ 2}
m^2+(1- z^{ 2})\kappa^2
 +{M \over 2} z~z' ({M \over 2} z_i + k^-_i ) +
 2 z'cos \theta \sqrt{\gamma
\gamma_i}-i\epsilon]^2} \ee
where $ z= -2 k^+/M$, with $1\geq|z|$  (from the
request of dealing with particles only, i.e. $p^+/2 \pm k^+\geq 0$ in the
valence wave-function).

It should be pointed out that the new term, $k\cdot k_i$, in the
denominator of Eq. (\ref{ptirsc}) does not  produce substantial differences in the formal
treatment that we applied to the bound-state case (cf Eq. (\ref{bsint})).
Therefore,
performing analogous  formal steps, one
obtains
 an inhomogeneous
 integral equation for the Nakanishi weight
function, $g^{(+)}(\gamma,z,z';\gamma_i,z_i)$. In particular, the inhomogeneous
 integral equation is obtained by
inserting (\ref{bsintscat}) in both sides of the following 4D equation,
obtained from Eq. (\ref{bs})\be
\left[\Phi^{(+)}(k,p)-(2\pi)^4\delta^{(4)}(k-k_i)\right ]= G_0^{(12)}(k,p)
~
i{\cal K}(k,k_i,p)+ \nonu +G_0^{(12)}(k,p)
~
\int
\frac{d^4k^\prime}{(2\pi)^4}i~{\cal K}(k,k^\prime,p)\left[\Phi^{(+)}(k^\prime,p)-
(2\pi)^4\delta^{(4)}(k^\prime-k_i)\right ]
\label{bsecon}\ee
 and then by projecting
onto the LF plane, as shown in Eq. (\ref{BST}) for the bound states,
viz \be \int {dk^-\over 2 \pi}
\left[\Phi^{(+)}(k,p)-(2\pi)^4\delta^{(4)}(k-k_i)\right ]= \int
{dk^-\over 2 \pi}G_0^{(12)}(k,p) ~i{\cal K}(k,k_i,p)+ \nonu + \int
{dk^-\over 2 \pi}G_0^{(12)}(k,p) ~ \int
\frac{d^4k^\prime}{(2\pi)^4}i~{\cal K}(k,k^\prime,p)
\left[\Phi^{(+)}(k^\prime,p)-
(2\pi)^4\delta^{(4)}(k^\prime-k_i)\right ] \label{lfbsec}\ee
After introducing  Eq. (\ref{ptirsc}) in Eq. (\ref{lfbsec}) and
integrating over $k^-$,
 one obtains the following integral equation for
the Nakanishi
amplitude, $g^{(+)}(\gamma,z',z;\gamma_i,z_i)$, 
 without angular momentum decomposition, viz
\be \label{bsnewscatt}
\int_{-1}^1 dz'\int_{-\infty}^{\infty}d\gamma'
\frac{g^{(+)}(\gamma',z',z;\gamma_i,z_i)}
{[\gamma'+\gamma+ z^{ 2}
m^2+(1- z^{ 2})\kappa^2
 +{M \over 2} z~z' ({M \over 2} z_i + k^-_i ) +
 2 z'cos \theta \sqrt{\gamma
\gamma_i}-i\epsilon]^2}=\nonu= {\cal
I}^{LF}(\gamma,z;\gamma_i,z_i,cos \theta) +\nonu +
\int_{-\infty}^{\infty}d\gamma'\int_{-1}^{1}d\zeta\int_{-1}^{1}d\zeta'
~ V^{LF}_s(\gamma,z;\gamma_i,z_i,\gamma',\zeta,\zeta',cos \theta)
~g^{(+)}(\gamma',\zeta,\zeta';\gamma_i,z_i). \ee where the
inhomogeneous term is given by
\be
\label{inlf}
{\cal I}^{LF}(\gamma,z;\gamma_i,z_i,cos \theta)= p^+~\int {dk^-\over 2
\pi}G_0^{(12)}(k,p) ~i{\cal K}(k,k_i,p)=\nonu =-p^+~\int {dk^-\over
2 \pi} {1 \over \left[(\frac{p}{2}+ k)^2-m^2+i\epsilon\right]}~ {1
\over\left[(\frac{p}{2}-k)^2-m^2+i\epsilon\right]} ~i{\cal
K}(k,k_i,p) \ee
and the kernel $V^{LF}_s$ is related to the kernel $i{\cal
K}$ of  the BS equation by (cf Eq. (\ref{vbou}) for the bound state)
\be
\label{VS}
 V^{LF}_s(\gamma,z;;\gamma_i,z_i,\gamma',\zeta,\zeta',cos\theta)=~i
p^+~\int_{-\infty}^{\infty}{d k^- \over 2\pi}~G_0^{(12)}(k,p)
~\times \nonu\int \frac{d^4k''}{(2\pi)^4}\frac{i{\cal K}(k,k'',p)}
{\left[k^{\prime\prime 2}-\kappa^2 +p\cdot k''~\zeta^\prime
 +2 k''\cdot k_i~\zeta -\gamma' +i\epsilon \right ]^3}
\ee
 It is worth noting that, {\em modulo the value of $\kappa^2$}, one
formally has for the S-wave
\be
\lim_{\zeta \to 0}V^{LF}_s(\gamma,z;\gamma_i,z_i,\gamma',\zeta,\zeta',cos\theta)=V^{LF}_b(\gamma,z;\gamma',\zeta')
\ee
Notice that the dependence upon $\gamma_i$ and  $z_i$ is washed out by putting $\zeta=0$.
 (see subsct. \ref{vbrevis} for a discussion in ladder approximation).
 
 \subsection{The  scattering amplitude}
\label{scat}
In order to complete the theoretical analysis, let us present the relation
between the scattering amplitude and the Nakanishi weight function. For the sake
of simplicity, the frame where the $z-axis$ is perpendicular to the scattering
plane has been chosen, then the
 scattering amplitude  $f(s,\theta)$ (with $s=M^2$),
 can be   calculated from the BS
amplitude. As a matter of fact, one has (cf \cite{grossbook})
\be
f(s,\theta)= {1\over k_r}~\sum_{\ell}
(2\ell+1)~ e^{i \delta_\ell} ~sin\delta_\ell~P_\ell(cos \theta)
=\nonu=-{1\over M~8 \pi}~\lim_{k'\to k_f}~\langle p'_1,p'_2|~i{\cal K}(p)
|\Phi^{(+)};p,k_i\rangle=\nonu=-{1 \over M~8 \pi}~
\lim_{k'\to k_f}~\langle k',p|G_0^{-1}(p)|\Phi^{(+)};p,k_i\rangle
\ee
where $k_r=\sqrt{s/4-m^2}$, $k'=(p'_1-p'_2)/2=$, $p'_1+p'_2=p$ and the orthogonality of the plane waves
 has been adopted (cf Eq. (\ref{bs})). By using the LF projection method
 (Refs.\cite{sales00,sales01,adnei08}), one can  move  from the 4D Minkowski
 space
 to the
 3D LF hyperplane, simplifying the analytic integration, without approximations.
 Then, one can rewrite the scattering
 amplitude in terms of the LF 3D scattering wave function as follows
 \be f(s,\theta)=-{1 \over M~8 \pi}
 \lim_{k'\to k_f}\langle k',p|G_0^{-1}(p)|\Phi^{(+)};p,k_i\rangle=\nonu=
 -{1 \over M~8 \pi}~
 \lim_{\tilde k'\to \tilde k_f}\langle
 \tilde k'|\bar \Pi_0(p)~G_0^{-1}(p)~\Pi(p)
 |\phi^{(+)}_{LF};p,\tilde k_i \rangle
 \ee
 where, from Eqs. (\ref{psitolfa}) and (\ref{psitolfr}), one has
 \be
 |p,k' \rangle = \Pi_0~|\tilde k' \rangle ~~~,\quad \quad
 |\Phi^{(+)};p,k_i\rangle=
\Pi(p)~|\phi^{(+)}_{LF};p,\tilde k_i\rangle
 \ee
 Recalling that (see Eqs. (\ref{3Df})  and  (\ref{invop}), and Ref.
  \cite{adnei08})
 \be
 \bar \Pi_0(p)~G_0^{-1}(p)~\Pi(p)=\Pi_0(p)~G_0^{-1}(p)~\left [1 +\Delta_0(p)~W(p)~\right ]~ \Pi_{0}(p)
 \nonu=\bar \Pi_0(p)~G_0^{-1}(p)~\Pi_0(p)=
 g_0^{-1}(p)
 \ee
 since \be \overline \Pi_0(p)
~\inv{G_0(p)}\Delta_0(p)=\overline \Pi_0(p)-\overline \Pi_0(p)
~\inv{G_0(p)} \Pi_0(p) g_0(p) \overline \Pi_0(p)=0 ~~~~,
\label{delta}\ee
 one can  write
 \be
  f(s,\theta)=
 -{1 \over M~8 \pi}~~
 \lim_{\tilde k'\to \tilde k_f}\langle \tilde k'|g_0^{-1}(p)
 |\phi^{(+)}_{LF};p,\tilde k_i \rangle
 \ee
 Finally, by inserting Eq. (\ref{g0m2}) one gets
 \be f(s,\theta)=~{i \over M~8\pi}~
\lim_{(\gamma,z) \to (\gamma_f,z_f)}{p^+\over 4} (1-z^2)\left(
M^2-4\frac{m^2+\gamma}{1-z^2}\right)~
\phi^{(+)}_{LF}\left(z,{\gamma},cos\theta\right)=\nonu=
{i \over M~8\pi}~
\lim_{(\gamma,z) \to (\gamma_f,z_f)}\left(
M^2-4\frac{m^2+\gamma}{1-z^2}\right)~
\psi^{(+)}\left(z,{\gamma},cos\theta\right)
\ , \label{f0}
 \ee
 where  $\psi^{(+)}$ is given by Eq. (\ref{scat1}) and the
  dependence upon
  ${\gamma_i}=|{\bf k}_{i\perp}|^2$ is understood.

Equation (\ref{f0}) can be written in  terms of the Nakanishi representation
 by using
 i) the distorted part of $\psi^{(+)}$ and ii)
  Eq. (\ref{bsnewscatt}), viz
\be f(s,\theta)=~-~{i \over M~8~\pi}~\lim_{(\gamma,z) \to (\gamma_f,z_f)}
\left[\gamma +(1-z^2)\kappa^2 +z^2m^2
\right]~\left [
{\phantom\int} \! {\cal I}^{LF}(\gamma,z;\gamma_i,z_i)
+ \right. \nonu + \left.
\int_{-\infty}^{\infty}d\gamma'\int_{-1}^{1}d\zeta\int_{-1}^{1}d\zeta'\;
V^{LF}_s(\gamma,z;\gamma_i,z_i,\gamma',\zeta,\zeta',cos \theta)
g^{(+)}(\gamma',\zeta,\zeta';\gamma_i,z_i)\right] \ , \label{f0f}
 \ee
where  $\gamma_f=\gamma_i$ and
$z_f=z_i$.
Notice that the factor $\gamma +(1-z^2)\kappa^2 +z^2m^2 $
vanishing for $(\gamma,z)\to (\gamma_f,z_f)$, is canceled
out by the corresponding one in
${\cal I}^{LF}$ and $V^{LF}_s$. This will be illustrated in the next section
within the
ladder approximation.
\section{Scattering states in Ladder Approximation}
\label{seclad}
In this Section, we present the ladder approximation of the integral equation
(\ref{bsnewscatt}), in order to determine the corresponding Nakanishi amplitude,
$g^{(+)}_L(\gamma',z',z;\gamma_i,z_i)$. We also illustrate how uniqueness can be
explicitly exploited in order to get a simpler integral equation, but with a more
elaborated kernel.

\subsection{ The Nakanishi weight function integral equation}
In ladder approximation
(starting from now, we drop the superscript $LF$ to simplify the
notation), Eq. (\ref{bsnewscatt}) reads
\be \label{scatlad}
\int_{-1}^1 dz'\int_{-\infty}^{\infty}d\gamma'
\frac{g^{(+)}_L(\gamma',z',z;\gamma_i,z_i)}
{[\gamma'+\gamma+ z^{ 2}
m^2+(1- z^{ 2})\kappa^2
 +{M \over 2} z~z' ({M \over 2} z_i + k^-_i ) +
 2 z'cos \theta \sqrt{\gamma
\gamma_i}-i\epsilon]^2}=\nonu= {\cal I}^{(L)}(\gamma,z;\gamma_i,z_i,cos \theta) +\nonu +
\int_{-\infty}^{\infty}d\gamma''\int_{-1}^{1}d\zeta\int_{-1}^{1}d\zeta'
~ V^{(L)}_s(\gamma,z;\gamma_i,z_i,\gamma'',\zeta,\zeta',cos \theta)
~g^{(+)}(\gamma'',\zeta,\zeta';\gamma_i,z_i)~~. \ee
where ${\cal I}^{(L)}$ is given by (see  Appendix \ref{inho} for details)
\be
{\cal I}^{(L)}(\gamma,z;\gamma_i,z_i,cos \theta)=g^2 ~
{1 \over \left[\gamma +(1-z^2) \kappa^2 +z^2m^2 -i\epsilon \right]}
{\cal G}^{(L)}(\gamma,z;\gamma_i,z_i,cos \theta)
\ee
with
\be
{\cal G}^{(L)}(\gamma,z;\gamma_i,z_i,cos \theta)=~
{\theta(z-z_i) (1-z)\over  \beta(z,z_i)+\gamma(1-z_i) - 2(1-z)cos\theta~\sqrt{\gamma
\gamma_i}
  -i\epsilon}
+
\nonu +
 { \theta(z_i-z)(1+z)\over  \beta(-z,-z_i)+\gamma (1+z_i)-
2(1+z)cos\theta~\sqrt{\gamma \gamma_i}
   -i\epsilon}
\label{calgl}\ee
and
\be
\beta(z,z_i)=
(1-z)\left[\mu^2+{M^2\over
4}(1-z)(1+z_i) -2m^2\right ]+(1-z_i)m^2 ~~~.
\ee

In Eq. (\ref{scatlad}), $V^{(L)}_s$
(cf Appendix \ref{vlad} for details) is given by
\be
V^{(L)}_{s}(\gamma,z;\gamma_i,z_i,\gamma'',\zeta,\zeta',cos\theta)=
~{g^2 \over 2(4\pi)^2}~{1 \over \left[\gamma +(1-z^2)\kappa^2 +z^2 m^2
-i\epsilon\right]}
~\times \nonu
\int_0^1  d v ~v^2~{\cal F}(v,\gamma,z;\gamma'',\zeta,\zeta',
cos\theta)\ee
where (dropping the dependence upon the external variables $z_i$, and
$\gamma_i$ for the
sake of simplicity)
\be
{\cal F}(v,\gamma,z;\gamma'',\zeta,\zeta',cos\theta)=
{\cal C}(v,\gamma,z;\gamma'',\zeta,\zeta',z_i,
cos\theta)+ {\cal C}(v,\gamma,-z;\gamma'',\zeta,-\zeta',-z_i,
cos\theta)
\nonu\label{calf0}\ee
with
\be {\cal C}(v,\gamma,z;\gamma'',\zeta,\zeta',z_i,
cos\theta)=
{(1+z)^2\over X^2(v,z_i,\zeta,\zeta')}~\times \nonu
{\theta (\zeta'-z-z_i\zeta) \over \left[
\gamma +z^2 m^2+\kappa^2(1-z^2)+
\Gamma(v,z,z_i,\zeta,\zeta',\gamma'')
+ Z(z,\zeta,\zeta';z_i)
\left[{M^2\over 2} z z_i + 2  cos\theta \sqrt{\gamma \gamma_i}\right]
  -i\epsilon\right]^2}
\nonu\label{calfl}\ee
In Eq. (\ref{calfl}), one has
\be
X(v,z_i,\zeta,\zeta')=v(1-v)(1+\zeta'-z_i\zeta)
\nonu
\Gamma(v,z,z_i,\zeta,\zeta',\gamma'')=
{ (1+z)\over (1+\zeta'-z_i\zeta)}~\left\{{v\over (1-v)}\left[{\zeta'}^{2}
\frac{M^2}{4} + \kappa^2 (1+{\zeta}^2)+\gamma''\right]+{\mu^2\over v}
+\gamma''\right\} \nonu
Z(z,\zeta,\zeta';z_i)=\zeta~{ (1+z)\over (1+\zeta'-z_i\zeta)}\nonu
\label{xgz}\ee

Collecting the above results, one obtains the following integral
equations for the Nakanishi amplitude in ladder approximation, $g^{(+)}_L$,
\be
\label{ladscat}
\int_{-1}^1 dz'\int_{-\infty}^{\infty}d\gamma'
\frac{g^{(+)}_L(\gamma',z',z;\gamma_i,z_i)}
{[\gamma'+\gamma+ z^{ 2}
m^2+(1- z^{ 2})\kappa^2
   +{M^2 \over 2} z~z'  z_i+
 2 z'cos \theta \sqrt{\gamma
\gamma_i}-i\epsilon]^2}
 =\nonu=~g^2 ~ {1 \over \left[\gamma +(1-z^2) \kappa^2 +z^2m^2 -i\epsilon \right]}
{\cal G}^{(L)}(\gamma,z;\gamma_i,z_i,cos \theta)+\nonu
~+{g^2 \over 2(4\pi)^2}~
{1 \over \left[\gamma +(1-z^2) \kappa^2 +z^2m^2 -i\epsilon \right]}
\int_{-\infty}^{\infty}d\gamma''\int_{-1}^{1}d\zeta\int_{-1}^{1}d\zeta' ~
\times \nonu
\int_0^1  d v ~v^2{\cal F}(v,\gamma,z;\gamma'',\zeta,\zeta',
cos\theta)
~g^{(+)}_L(\gamma'',\zeta,\zeta';\gamma_i,z_i)
\ee
 A solution of Eq. (\ref{ladscat}), obtained by retaining only the
inhomogeneous term and assuming the uniqueness of the solution, is
discussed in the following Section. Such a solution can yield some insights
 on the
 the analytic structure of the
ladder approximation of the equation for determining the Nakanishi
amplitude $g^{(+)}_L$, i.e. Eq. (\ref{ladscat}). In particular,
one could argue that a finite lower
bound for the variable $\gamma'$ could exist in order to reproduce
the analytic structure of the inhomogeneous term.
In particular, the inhomogeneous term
is proportional to  the global propagator $$ {1 \over M^2
-{4(m^2+\gamma)\over (1-z^2)}+i\epsilon}=-
{1 \over \gamma+ (1 -z^2) \kappa^2 +z^2
m^2 -i\epsilon}$$ that generates a   cut starting at
$\gamma=-(1 -z^2) \kappa^2 -z^2m^2$. The lowest value is
$\gamma_{min}=-m^2\leq 0$ recalling that $z^2\in[0,1]$.
Therefore, since $\gamma=|{\bf
k}_\perp|^2$, the lowest bound is zero.

To conclude this subsection, it is shown
 the scattering amplitude, Eq. (\ref{f0f}), in ladder
approximation,viz

 \be f^{(L)}(s,\theta)= ~-i{g^2\over M 8\pi} ~
 \left[ {\cal G}^{(L)}(\gamma_f,z_f;\gamma_i,z_i,cos
\theta) ~+{1 \over 2^5\pi^{2}}~
\int_{-\infty}^{\infty}d\gamma''\int_{-1}^{1}d\zeta\int_{-1}^{1}d\zeta'
~ \times \right. \nonu \left. \int_0^1  d v ~v^2
{\cal F}(v,\gamma_f,z_f;\gamma'',\zeta,\zeta', cos\theta)
~g^{(+)}_L(\gamma'',\zeta,\zeta';\gamma_i,z_i)\right]
\label{f0lad} \ee
where ${\cal G}^{(L)}$ and  ${\cal F}$ are  given in Eqs. (\ref{calgl}) and (\ref{calf0}), respectively.

\subsection{Applying uniqueness to the integral equation}
Eq. (\ref{ladscat}) can be rewritten in such a way that the denominator
appearing in the lhs can be present in both the terms in the rhs. This
allow to  explicitly use the uniqueness of the solution of the integral
equation and to obtain a simpler equation, as shown in what follows.

By using the standard Feynman trick, one can  rewrite ${\cal I}^{(L)}$ in a
 useful form for the following elaboration, namely (see  Appendix \ref{inho} for a
 detailed discussion)
 \be
 {\cal I}^{(L)}(\gamma,z;\gamma_i,z_i,cos \theta)=
 g^2 ~ \int_{-1}^1 dz'~\theta(-z')~\int_{-\infty}^{\infty}d\gamma' ~\times \nonu
~{\delta(\gamma'-\gamma_a(z'))\over \left[
 \gamma'+  \gamma +(1-z^2)\kappa^2 +z^2 m^2+{M^2 \over 2} z~z'  z_i
 + 2z' cos\theta~\sqrt{\gamma
\gamma_i}-i\epsilon
\right]^2}~\times \nonu
\left \{ \theta(z-z_i)~\theta
[1-z+z'(1-z_i)]+  \theta(z_i-z)~\theta[1+z+z'(1+z_i)]  \right\}
\label{ilbody} \ee
 where \be
\gamma_a(z')
=z'\left(2\kappa^2-\mu^2\right )
\label{gammaa}\ee

As for  $V^{(L)}_s$, one can proceed through more subtle mathematical steps.
  This is thoroughly discussed   in
Appendix \ref{vlad}, here  the final expression is given. One has
\be
V^{(L)}_{s}(\gamma,z;\gamma_i,z_i,\gamma'',\zeta,\zeta',cos\theta)= ~-
~{g^2 \over 2(4\pi)^2}~\times \nonu \int^\infty_{-\infty}
 d\gamma' \int_{-1}^1 {dz'}{1 \over  \left[\gamma +z^2 m^2+
\kappa^2(1-z^2)+\gamma'+z'\left({M^2\over 2} z z_i + 2  cos\theta \sqrt{\gamma
\gamma_i}\right)
-i\epsilon   \right]^2}
 \times \nonu \left [{(1+z)\over (1+\zeta'-z_i\zeta)}
~\theta (\zeta'-z-z_i\zeta)~
{\cal Q}'(z,z_i;\gamma'',\gamma',z',\zeta,\zeta',\mu^2)
+\right. \nonu \left . +{(1-z)\over (1-\zeta'+z_i\zeta)}
~\theta
(z-\zeta'+z_i\zeta)~
{\cal Q}'(-z,-z_i;\gamma'',\gamma',z',\zeta,-\zeta',\mu^2)\right]
\label{vlsbody}
\ee
where
\be
{\cal Q}'(z,z_i;\gamma'',\gamma',z',\zeta,\zeta',\mu^2)=
\theta\left({1+z \over 1+\zeta' -z_i \zeta} -{z'\zeta}\right)
{ \theta(z')~\theta(\zeta) - \theta(-z')~ \theta(-\zeta)
 \over z'}~\times \nonu
\Lambda\left(z,{z'\over \zeta},\zeta,\zeta';\gamma'',\gamma';z_i,\mu^2\right)
\label{qprime}\ee 
 where
 \be
\Lambda\left(z,{z'\over \zeta},\zeta,\zeta';\gamma'',\gamma';z_i,\mu^2\right)
= 
\sum_{i=\pm} {\partial \over \partial \lambda} y_i(0)\left\{ \delta(y_i(0)) ~
{y^2_i(0) \over |y^2_i(0){\cal A}(\zeta,\zeta',\gamma'',\kappa^2)
 -\mu^2|}+ \right . \nonu \left.
  -\theta(y_i(0)) ~ { 2\mu^2~ y_i(0) \over (y^2_i(0){\cal
  A}(\zeta,\zeta',\gamma'',\kappa^2)
 -\mu^2)|y^2_i(0){\cal A}(\zeta,\zeta',\gamma'',\kappa^2)
 -\mu^2|}\right\}
 \label{qprime1}\ee
 with $y_\pm(0)$ solutions of the following second order equation
 \be
 y^2~{\cal A}(\zeta,\zeta',\gamma'',\kappa^2)  +y~
 {\cal B}(z,z',z_i,\zeta,\zeta',\gamma'',\gamma',\mu^2,\lambda) +\mu^2=0
 \label{seceq}\ee
 The dependence upon $z'/\zeta~\geq~ 0$ will become clear from what follows.
 
 In Eq. (\ref{seceq}) the coefficients are given by
 \be
{\cal A}(\zeta,\zeta',\gamma'',\kappa^2)={\zeta'}^{2} \frac{M^2}{4} +
\kappa^2 (1+{\zeta}^2)+\gamma''
\nonu
{\cal B}(z,z',z_i,\zeta,\zeta',\gamma'',\gamma',\mu^2,\lambda)=\mu^2 +\gamma'-
\gamma'' {\zeta \over z'} +\lambda {(1+\zeta'-z_i\zeta)\over (1+z)}
\label{seceq1}\ee
In Eq. (\ref{qprime}), one explicitly has
 \be
y_\pm(0)=
{1 \over 2{\cal A}(\zeta,\zeta',\gamma'',\kappa^2)} ~\times \nonu
 \left[ -{\cal B}(z,z',z_i,\zeta,\zeta',\gamma'',\gamma',\mu^2,0)
 \pm 
\sqrt{{\cal B}^2(z,z',z_i,\zeta,\zeta',\gamma'',\gamma',\mu^2,0)
- 4 \mu^2~ {\cal
A}(\zeta,\zeta',\gamma'',\kappa^2)} \right]
\nonu
{\partial \over \partial \lambda} y_i(0)=~\mp~
{(1+\zeta'-z_i\zeta)\over (1+z)}~
~{ y_\pm(0)\over
\sqrt{{\cal B}^2(z,z',z_i,\zeta,\zeta',\gamma'',\gamma',\mu^2,0)
- 4 \mu^2{\cal A}(\zeta,\zeta',\gamma'',\kappa^2)}}
\label{seceq2}\ee
Notice that for $\lambda=0$, the dependence upon $z_i$ in
${\cal B}(z,z',z_i,\zeta,\zeta',\gamma'',\gamma',\mu^2,0)$ is dummy.

Eq. (\ref{qprime}) can be simplified, noting that  the first term,
  proportional to
$y^2_i(0)~\delta(y_i(0))$,  does not contribute if $\mu^2 \ne 0$, while for
$\mu^2=0$ (Wick-Cutkosky model, see the next Section) the second term,
proportional
to $\theta(y_i(0))$, is vanishing.

By using Eq. (\ref{ilbody}) and  Eqs (\ref{vlsbody}), (\ref{ladscat}) can be rewritten putting in evidence the analytic behavior of all
the terms, namely
\be
\label{ladscatc}
\int_{-1}^1 dz'\int_{-\infty}^{\infty}d\gamma'
\frac{g^{(+)}_L(\gamma',z',z;\gamma_i,z_i)}
{[\gamma'+\gamma+ z^{ 2}
m^2+(1- z^{ 2})\kappa^2
   +{M^2 \over 2} z~z'  z_i+
 2 z'cos \theta \sqrt{\gamma
\gamma_i}-i\epsilon]^2}
 =\nonu=~g^2 ~ \int_{-1}^1 dz'~\theta(-z')~\int_{-\infty}^{\infty}d\gamma'
~{\delta[\gamma'-z'(2\kappa^2-\mu^2)]\over \left[
 \gamma'+  \gamma +(1-z^2)\kappa^2 +z^2 m^2+{M^2 \over 2} z~z'  z_i
 + 2z' cos\theta~\sqrt{\gamma
\gamma_i}-i\epsilon
\right]^2}~\times \nonu
\left\{ \theta(z-z_i)~\theta
[1-z+z'(1-z_i)]+  \theta(z_i-z)~\theta[1+z+z'(1+z_i)]  \right\}
+\nonu
~-{g^2 \over 2(4\pi)^2}~\int_{-1}^1 dz'~\int_{-\infty}^{\infty}d\gamma'
{1 \over  \left[\gamma +z^2 m^2+
\kappa^2(1-z^2)+\gamma'+z'\left({M^2\over 2} z z_i + 2  cos\theta \sqrt{\gamma
\gamma_i}\right)
-i\epsilon   \right]^2}
 \times \nonu \int_{-1}^1 d\zeta~\int_{-1}^1 d\zeta'~
 \int_{-\infty}^{\infty}d\gamma''\left [{(1+z)\over (1+\zeta'-z_i\zeta)}
~\theta (\zeta'-z-z_i\zeta)~
{\cal Q}'(z,z_i;\gamma'',\gamma',z',\zeta,\zeta',\mu^2)
+\right. \nonu \left . +{(1-z)\over (1-\zeta'+z_i\zeta)}
~\theta
(z-\zeta'+z_i\zeta)~
{\cal Q}'(-z,-z_i;\gamma'',\gamma',z',\zeta,-\zeta',\mu^2)\right]
~g^{(+)}_L(\gamma'',\zeta,\zeta';\gamma_i,z_i)
\ee
From the uniqueness of the solution of the integral equation (\ref{ladscatc}),
that we expect once  the Nakanishi theorem for the vertex function
 (cf \cite{nak63}) is extended to the scattering case,
one could write (see also \cite{nak642})
\be
g^{(+)}_L(\gamma',z',z;\gamma_i,z_i)
 =~g^2 ~ \theta(-z')~\delta[\gamma'-z'(2\kappa^2-\mu^2)]
\left\{ \theta(z-z_i)~\theta
[1-z+z'(1-z_i)]~
  +
 \right. \nonu
 \left.+  \theta(z_i-z)~\theta[1+z+z'(1+z_i)]  \right\}
~-{g^2 \over 2(4\pi)^2}~\int_{-\infty}^{\infty}d\gamma''\int_{-1}^1 d\zeta~
\int_{-1}^1 d\zeta'~g^{(+)}_L(\gamma'',\zeta,\zeta';\gamma_i,z_i)
 \times \nonu \left [{(1+z)\over (1+\zeta'-z_i\zeta)}
~\theta (\zeta'-z-z_i\zeta)~
{\cal Q}'(z,z_i;\gamma'',\gamma',z',\zeta,\zeta',\mu^2)
+\right. \nonu \left . +{(1-z)\over (1-\zeta'+z_i\zeta)}
~\theta
(z-\zeta'+z_i\zeta)~
{\cal Q}'(-z,-z_i;\gamma'',\gamma',z',\zeta,-\zeta',\mu^2)\right]
\label{gladder}\ee

The first term in the rhs of Eq. (\ref{gladder}) yields the lowest order
approximation to $g^{(+)}_L$. The following support for $0\leq \gamma'\leq \mu^2
+2 |\kappa^2|$ can be obtained by inspecting the delta function, combined with
the constraint on $z'$.

\section{Applications}
\label{applic}
In this Section,  some relevant limiting cases,  i) the Nakanishi amplitude for
scattering states at zero-energy ($\kappa^2\to 0$, $\mu^2\ne 0$),  ii) the
Nakanishi amplitude for the Wick-Cutkosky
model in the continuum  ($\mu^2 \to0$ and $\kappa^2\leq 0$),  and  iii) a
revisiting of the ladder kernel for bound states  are illustrated in details.
It is worth noting that at the end of the Wick-Cutkosky model subsection, it is
presented a formal comparison between the kernel for zero energy, obtained
within our approach and the kernel that one can find  in Ref. \cite{dae}, for a S-wave
bound state, putting a vanishing binding energy energy. It is rewarding to
find a successful comparison with
 our formalism.  Moreover, as shown in the last
 subsection, a new form of the integral equation that
 determines the Nakanishi amplitude for bound states, in ladder approximation,
 is obtained. This simple integral equation could suggest 
 a different numerical investigation of the issue.


\subsection{The Nakanishi Integral Equation for the zero-energy scattering}
\label{zeroen}
The choice $z_i=\gamma_i=0$ yields $\kappa^2=-\gamma_i -z^2_iM^2/4=0$,
namely a zero-energy scattering, given the physical meaning of
$\kappa^2=m^2-M^2/4$. For such values of $z_i$ and $\gamma_i$, one can
simplify Eq. (\ref{ladscatc}), obtaining the
Nakanishi integral equation for  the weight function at zero energy.
Notice that such a  weight amplitude determines the
scattering length  through Eq. (\ref{f0lad}). After inserting
$z_i=\gamma_i=\kappa^2=0$ in 
 Eq. (\ref{ladscatc}), one gets
\be
\int_{-1}^1 dz'\int_{-\infty}^{\infty}d\gamma'
\frac{g^{(+)}_L(\gamma',z',z;\gamma_i=z_i=0)}
{[\gamma'+\gamma+ z^{ 2}
m^2-i\epsilon]^2}
 =\nonu=~g^2 ~ \int_{-1}^1 dz'~\theta(-z')~\int_{-\infty}^{\infty}d\gamma'
~{\delta(\gamma'+z'\mu^2)\over \left[
 \gamma'+  \gamma  +z^2 m^2-i\epsilon
\right]^2}~\times \nonu
\left\{ \theta(z)~\theta
(1-z+z')~
  +  \theta(-z)~\theta(1+z+z')  \right\}
+\nonu
~-{g^2 \over 2(4\pi)^2}~\int_{-1}^1 dz'~\int_{-\infty}^{\infty}d\gamma'
{1 \over  \left[\gamma'+\gamma +z^2 m^2+
-i\epsilon   \right]^2}
 \times \nonu \int_{-1}^1 d\zeta~\int_{-1}^1 d\zeta'~\int_{-\infty}^{\infty}
 d\gamma''\left [{(1+z)\over (1+\zeta')}
~\theta (\zeta'-z)~
{\cal Q}'(z,z_i=0;\gamma'',\gamma',z',\zeta,\zeta',\mu^2)
+\right. \nonu \left . +{(1-z)\over (1-\zeta')}
~\theta
(z-\zeta')~
{\cal Q}'(-z,z_i=0;\gamma'',\gamma',z',\zeta,-\zeta',\mu^2)\right]
~g^{(+)}_L(\gamma'',\zeta,\zeta';\gamma_i=z_i=0)
\label{zeroeq}\ee
One immediately realizes that the integration over $z'$ can be performed in the
lhs, introducing the suitable S-wave Nakanishi amplitude for the present case.
Then, one
gets
\be
\int_{-\infty}^{\infty}d\gamma'
\frac{g^{(+)}_{0L}(\gamma',z)}
{[\gamma'+\gamma+ z^{ 2}
m^2-i\epsilon]^2}
 =\nonu=
{g^2\over \mu^2}  ~\int_{-\infty}^{\infty}d\gamma'
~{\theta(\gamma')\over
\left[ \gamma'+ \gamma +z^2 m^2
 -i\epsilon\right]^2 }~
\left\{ \theta(z)~\theta(1-z-\gamma'/\mu^2)
 +  \theta(-z)~\theta(1+z-\gamma'/\mu^2)  \right\}
+\nonu
~-{g^2 \over 2(4\pi)^2}~\int_{-\infty}^{\infty}d\gamma'
{1 \over  \left[\gamma' +\gamma +z^2 m^2+
-i\epsilon   \right]^2}
 \times \nonu ~\int_{-1}^1 d\zeta'~\int_{-\infty}^{\infty}
 d\gamma''\left [{(1+z)\over (1+\zeta')}
~\theta (\zeta'-z)~
{\cal T}(z,\gamma'',\gamma',\zeta',\mu^2)
+\right. \nonu \left . +{(1-z)\over (1-\zeta')}
~\theta
(z-\zeta')~
{\cal T}(-z,\gamma'',\gamma',-\zeta',\mu^2)\right]
~g^{(+)}_{0L}(\gamma'',\zeta')
\label{zeroe}\ee
where
\be
g^{(+)}_{0L}(\gamma',z)= \int_{-1}^1
dz'~g^{(+)}_L(\gamma',z',z;\gamma_i=z_i=0)
\nonu
\int_{-1}^0 dz'~\delta(\gamma'+z'\mu^2)= {1 \over \mu^2}~\theta(\gamma')
\nonu
g^{(+)}_{0L}(\gamma'',\zeta')=\int_{-1}^1 d\zeta~
g^{(+)}_L(\gamma'',\zeta,\zeta';\gamma_i=z_i=0)
\ee
and the possibility to integrate over $\zeta$ the function 
$g^{(+)}_L(\gamma'',\zeta,\zeta';\gamma_i=z_i=0)$ is related to the independence
upon $\zeta$ of the function $${\cal T}(z,\gamma'',\gamma',\zeta',\mu^2)=\int_{-1}^1 dz'~
{\cal Q}'(z,z_i=0;\gamma'',\gamma',z',\zeta,\zeta',\mu^2)~~.$$
As a matter of fact, from Eq. (\ref{qprime}) 
one has\be
{\cal T}(z,\gamma'',\gamma',\zeta',\mu^2)=
\int_{-1}^1 {dz'\over z'}~\theta\left(  {1+z \over 1+\zeta'}-{z'\over
\zeta}\right)\left[ \theta(z')~\theta(\zeta)
 - \theta(-z')~\theta(-\zeta)
 \right]
\times 
 \nonu\Lambda_0\left(z,{z'\over \zeta},\zeta',\gamma',\gamma'',\mu^2\right)=
 \nonu=
\left\{\theta(\zeta)\int_0^{\zeta{(1+z) \over (1+\zeta')}} {dz'\over z'}~
 -\theta(-\zeta)\int^0_{\zeta{(1+z) \over (1+\zeta')}} {dz'\over
 z'}\right\}~\Lambda_0\left(z,{z'\over \zeta},\zeta',\gamma',\gamma'',\mu^2\right) 
\label{calt}\ee
where
\be
\Lambda_0\left(z,{z'\over \zeta},\zeta',\gamma',\gamma'',\mu^2\right)=
\sum_{i=\pm} {\partial \over \partial \lambda} y_i(0)
~{1 \over|y^2_i(0){\cal A}_0(\zeta',\gamma'')
 -\mu^2|} ~ \times \nonu\left[ \delta(y_i(0)) y^2_i(0)
  -\theta(y_i(0)) ~ { 2\mu^2~ y_i(0) \over y^2_i(0){\cal
  A}_0(\zeta',\gamma'')
 -\mu^2}\right]
 \ee
 with the dependence upon $z'/\zeta$ (always positive !) explained in what follows.
 From Eqs. (\ref{seceq1}) and (\ref{seceq2}), one can explicitly write
 \be
{\cal A}_0(\zeta',\gamma'')={\zeta'}^{2} \frac{M^2}{4} +
\gamma''={\zeta'}^{2} m^2 +
\gamma''
\nonu
{\cal B}(z,z',0,\zeta,\zeta',\gamma',\gamma'',\mu^2,0)=\mu^2 +\gamma''-
\gamma' {\zeta \over z'}
\nonu
y_\pm(0)=
{1 \over 2 {\cal A}_0(\zeta',\gamma'')} ~ \times \nonu
\left[ -\left(\mu^2 +\gamma''-
\gamma' {\zeta \over z'} \right)
\pm ~
\sqrt{\left(\mu^2 +\gamma''- \gamma' {\zeta \over z'}\right)^2
- 4 \mu^2{\cal A}_0(\zeta',\gamma'')}  \right]
\nonu
{\partial \over \partial \lambda} y_i(0)=~\mp~
{(1+\zeta')\over (1+z)} 
~{ y_\pm(0)\over \sqrt{ (\mu^2 +\gamma''-
\gamma' {\zeta \over z'})^2
- 4 \mu^2{\cal A}_0(\zeta',\gamma'')}} \ee
This allows one to single out the dependence upon $z'\over \zeta$ of the
integrand in Eq. (\ref{calt}) and,
after introducing the variable
$$ x= {z'\over \zeta}~{(1+\zeta')\over (1+z)}
$$
 one has
\be
{\cal T}(z,\gamma'',\gamma',\zeta',\mu^2)= \left[ \theta(\zeta)
+\theta(-\zeta)\right]~\int_0^1 {dx \over x}
\Lambda_0
\left(z,x {(1+z)\over(1+\zeta')} ,\zeta',\gamma',\gamma'',\mu^2\right)\label{calt1}\ee
This completes the proof that ${\cal T}$ does not depend upon
  $\zeta$. Finally, from uniqueness, one can rewrite the inhomogeneous
 integral equation (\ref{zeroe}) in a simpler form, viz
\be
g^{(+)}_{0L}(\gamma',z)
 =
{g^2\over \mu^2}  ~\theta(\gamma')
\left [ \theta(z)~\theta(1-z-\gamma'/\mu^2)
 +  \theta(-z)~\theta(1+z-\gamma'/\mu^2)  \right]
+\nonu
~-{g^2 \over 2(4\pi)^2}~
  ~\int_{-1}^1 d\zeta'~\int_{-\infty}^{\infty}
 d\gamma''\left [{(1+z)\over (1+\zeta')}
~\theta (\zeta'-z)~
{\cal T}(z,\gamma'',\gamma',\zeta',\mu^2)
+\right. \nonu \left . +{(1-z)\over (1-\zeta')}
~\theta
(z-\zeta')~
{\cal T}(-z,\gamma'',\gamma',-\zeta',\mu^2)\right]
~g^{(+)}_{0L}(\gamma'',\zeta')
\label{gzero2}\ee
Such an equation could have a direct application in the study of the
relativistic effects of the scattering length.


\subsection{The Nakanishi amplitude for the Wick-Cutkosky model in the continuum}
\label{WiCu}
The Wick-Cutkosky model \cite{WICK_54,CUT_54}, namely two massive
scalars interacting through a massless one ($\mu^2=0$) in ladder
approximation, can be extended
to the scattering case. The integral equation for the 
Nakanishi amplitude, for this widely adopted model, was known so far only 
for  the bound state case (cf \cite{WICK_54,CUT_54,nakanishi} and   \cite{dae}
for a LF approach).
Equation (\ref{gladder}) becomes
\be
g^{(+)}_{LW}(\gamma',z',z;\gamma_i,z_i)
 =~g^2 ~ \theta(-z')~\delta(\gamma'-2z'\kappa^2)
\left[ \theta(z-z_i)~\theta[1-z+z'(1-z_i)]~
  +
 \right. \nonu
 \left.+  \theta(z_i-z)~\theta[1+z+z'(1+z_i)]  \right]
+\nonu
~-{g^2 \over 2(4\pi)^2}~\int_{-\infty}^{\infty}d\gamma''\int_{-1}^1 d\zeta~
\int_{-1}^1 d\zeta'~g^{(+)}_{LW}(\gamma'',\zeta,\zeta';\gamma_i,z_i)
 \times \nonu \left [{(1+z)\over (1+\zeta'-z_i\zeta)}
~\theta (\zeta'-z-z_i\zeta)~
{\cal Q}'(z,z_i;\gamma'',\gamma',z',\zeta,\zeta',\mu^2=0)
+\right. \nonu \left . +{(1-z)\over (1-\zeta'+z_i\zeta)}
~\theta
(z-\zeta'+z_i\zeta)~
{\cal Q}'(-z,-z_i;\gamma'',\gamma',z',\zeta,-\zeta',\mu^2=0)\right]
\label{gWick}\ee
where
\be
{\cal Q}'(z,z_i;\gamma',\gamma'',z',\zeta,\zeta',\mu^2=0)=
\theta\left({1+z \over 1+\zeta'-z_i \zeta} -{z'\zeta}\right) 
{ \theta(z')~\theta(\zeta) - \theta(-z')~\theta(-\zeta)
 \over z'}~\times \nonu
 {\partial \over \partial \lambda} y_0(0)~ \delta(y_0(0)) ~
{1\over |{\cal A}(\zeta,\zeta',\gamma'')|}
 \label{qWC}\ee
 with $y_0(0)$  being the non trivial the solution of the following second order equation
 \be
{\cal A}(\zeta,\zeta',\gamma'') y^2 +y
 {\cal B}(z,z',z_i,\zeta,\zeta',\gamma'',\mu^2=0,\lambda)=0
 \label{seceqW}\ee
Notice that the trivial solution, $y=0$ for any $\lambda$, does not 
contribute, due to the presence of ${\partial y_0(0)/\partial \lambda}$.
 In Eq. (\ref{seceqW}) the coefficients are given by
 \be
{\cal A}(\zeta,\zeta',\gamma'')={\zeta'}^{2} \frac{M^2}{4} +
\kappa^2 (1+{\zeta}^2)+\gamma''
\nonu
{\cal B}(z,z',z_i,\zeta,\zeta',\gamma'',\mu^2=0,\lambda)=\gamma''-
\gamma' {\zeta \over z'} +\lambda {(1+\zeta'-z_i\zeta)\over (1+z)}
\label{seceqW1}\ee
Then, one explicitly has
 \be
y_0(0)= { \gamma' {\zeta \over z'}-\gamma''\over {\cal A}(\zeta,\zeta',\gamma'',\kappa^2)}
\nonu
{\partial \over \partial \lambda} y_0(0)=~-~
{(1+\zeta'-z_i\zeta)\over (1+z)}
{ 1 \over{\cal A}(\zeta,\zeta',\gamma'',\kappa^2)}
\label{seceqW2}\ee
It has to be pointed out that, if  ${\cal A}(\zeta,\zeta',\gamma'')=0$,
one has only
the trivial solution $y=0$ for any $\lambda$.

 Equation (\ref{gWick}) can be rewritten as follows
\be
g^{(+)}_{LW}(\gamma',z',z;\gamma_i,z_i)
 =
~g^2 ~ \theta(-z')~\delta(\gamma'-2z'\kappa^2)
\left[ \theta(z-z_i)~\theta[1-z+z'(1-z_i)]~
  +
 \right. \nonu
 \left.+  \theta(z_i-z)~\theta[1+z+z'(1+z_i)]  \right]
+\nonu
~+{g^2 \over 2(4\pi)^2}~ \theta(z')~
 \int_{-\infty}^{\infty}d\gamma''\int_{0}^1 {d\zeta\over
z'}~
\int_{-1}^1 d\zeta'~
{\delta\left( \gamma' {\zeta \over z'}-\gamma''\right)\over
{\cal A}(\zeta,\zeta',\gamma'',\kappa^2)}
~g^{(+)}_{LW}(\gamma'',\zeta,\zeta';\gamma_i,z_i)
 \times \nonu \left \{
~\theta (\zeta'-z-z_i\zeta)~
\theta \left({ 1+z\over 1+\zeta'-z_i\zeta}- {z'\over \zeta}\right)
+ 
~\theta(z-\zeta'+z_i\zeta)~
\theta\left({ 1-z\over 1-\zeta'+z_i\zeta} -{z'\over \zeta}\right)
 \right\}
 \nonu -{g^2 \over 2(4\pi)^2} \theta(-z')~
 \int_{-\infty}^{\infty}d\gamma''\int_{-1}^0 {d\zeta\over
z'}~
\int_{-1}^1 d\zeta'~
{\delta\left( \gamma' {\zeta \over z'}-\gamma''\right)\over
{\cal A}(\zeta,\zeta',\gamma'',\kappa^2)}
~g^{(+)}_{LW}(\gamma'',\zeta,\zeta';\gamma_i,z_i)
 \times \nonu 
 \left \{
~\theta (\zeta'-z-z_i\zeta)~
\theta \left({ 1+z\over 1+\zeta'-z_i\zeta}- {z'\over \zeta}\right)
+ 
~\theta(z-\zeta'+z_i\zeta)~
\theta\left({ 1-z\over 1-\zeta'+z_i\zeta} -{z'\over \zeta}\right)
 \right\}\label{gWick2}
 \ee
 Notice that the extrema of the integration on $\zeta$ in the second and third
 terms are different.
 It is easily seen that, in the rhs,
  the positive $z'$ decouples from the negative one, obtaining a homogeneous
  integral equation for $\theta(z')~g^{(+)}_{LW}(\gamma',z',z;\gamma_i,z_i)$ and
   an inhomogeneous one for
   $\theta(-z')~g^{(+)}_{LW}(\gamma',z',z;\gamma_i,z_i)$.
   Therefore, the Nakanishi amplitude for scattering states within the
   Wick-Cutkosky model is given by the following inhomogeneous integral equation
   (it is inhomogeneous for matching the boundary condition for scattering
   states)
   \be
   g^{(+)}_{LW}(\gamma',z',z;\gamma_i,z_i)=
~g^2 ~ \theta(-z')~\delta(\gamma'-2z'\kappa^2)
\left[ \theta(z-z_i)~\theta[1-z+z'(1-z_i)]~
  +
 \right. \nonu
 \left.+  \theta(z_i-z)~\theta[1+z+z'(1+z_i)]  \right]
+\nonu
-{g^2 \over 2(4\pi)^2} \theta(-z')~
 \int_{-\infty}^{\infty}d\gamma''\int_{-1}^0 {d\zeta\over
z'}~
\int_{-1}^1 d\zeta'~
{\delta\left( \gamma' {\zeta \over z'}-\gamma''\right)\over
{\cal A}(\zeta,\zeta',\gamma'',\kappa^2)}
~g^{(+)}_{LW}(\gamma'',\zeta,\zeta';\gamma_i,z_i)
 \times \nonu 
 \left \{
~\theta (\zeta'-z-z_i\zeta)~
\theta \left({ 1+z\over 1+\zeta'-z_i\zeta}- {z'\over \zeta}\right)
+ 
\theta(z-\zeta'+z_i\zeta)~
\theta\left({ 1-z\over 1-\zeta'+z_i\zeta} -{z'\over \zeta}\right)
 \right\}
  \label{gWick3} \ee
   Let us assume the following separable form for
   $$g^{(+)}_{LW}(\gamma',z',z;\gamma_i,z_i)=
   \theta(-z')~\delta(\gamma'-2z'\kappa^2)~ h^{(+)}_{LW}(z',z;\gamma_i,z_i)~~.
   $$ Then, by integrating both sides of Eq. (\ref{gWick3})
   on $\gamma'$,
   one has (recalling that $\zeta/z'\geq 0$)
   \be
  h^{(+)}_{LW}(z',z;\gamma_i,z_i)=
    g^2 ~
\left\{ \theta(z-z_i)~\theta[1-z+z'(1-z_i)]~
  +
   \theta(z_i-z)~\theta[1+z+z'(1+z_i)]  \right\}
+\nonu
-{g^2 \over 2(4\pi)^2}
 \int_{-1}^0 {d\zeta\over
\zeta}~
\int_{-1}^1 d\zeta'~
{1\over{\zeta'}^{2} \frac{M^2}{4} +
\kappa^2 (1+{\zeta})^2}
~h^{(+)}_{LW}(\zeta,\zeta';\gamma_i,z_i)
 \times \nonu 
\left\{
~\theta (\zeta'-z-z_i\zeta)~
\theta \left({ 1+z\over 1+\zeta'-z_i\zeta}- {z'\over \zeta}\right)
+ 
\theta(z-\zeta'+z_i\zeta)~
\theta\left({ 1-z\over 1-\zeta'+z_i\zeta} -{z'\over \zeta}\right)
 \right\}
  \label{gWick4} \ee
   It is worth noting that the solution $h^{(+)}_{LW}$
   is perfectly compatible with the separable form
    we assumed, and that ${\zeta'}^{2} \frac{M^2}{4} +
\kappa^2 (1+{\zeta})^2= {\cal A}(\zeta,\zeta',2\zeta\kappa^2,\kappa^2)$ is
different from zero, as discussed below Eq. (\ref{seceqW2}).

Summarizing, the solution of Eq. (\ref{gWick4}), together with the separable form
mentioned above, allows one to obtain the Nakanishi amplitude for scattering
states within the Wick-Cutkosky model in the continuum.

As a formal check of our previous elaboration, in what follows we devise  
the same kernel, one has for the Wick-Cutkosky model for bound states in
the limit $\kappa^2\to 0$. This allows us to make a successful comparison with
 the result  for the S-wave, in the limit of zero-binding energy, that one can get
from  Ref. \cite{dae}.

In the limit $$\kappa^2=m^2 -{M^2\over 4}=-\gamma_i-z^2_i{M^2\over 4} \to 0$$
Eq. (\ref{gWick4}) becomes (dropping the dependence upon $z_i=\gamma_i=0$)
\be
  h^{(+)}_{LW}(z',z)=
    g^2 ~
\left[ \theta(z)~\theta[1-z+z']
+  \theta(-z)~\theta[1+z+z']  \right]
+\nonu
-{g^2 \over 2(4\pi)^2}
 \int_{-1}^0 {d\zeta\over
\zeta}~
\int_{-1}^1 d\zeta'~
{1\over{\zeta'}^{2} m^2 }
~h^{(+)}_{LW}(\zeta,\zeta')
 \times \nonu \left \{
~\theta (\zeta'-z)~
\theta[z'-\zeta { (1+z)\over (1+\zeta')} ]
 +
~\theta(z-\zeta')~
\theta[z'-\zeta{ (1-z)\over (1-\zeta')}  ]
 \right\}
  \label{gWick5} \ee
By integrating over $z'$ (let us recall that $0\geq z'\geq -1$), 
in order to match the S-wave dependence, one can write

\be
 a^{(+)}_{LW}(z) =
    g^2 ~
\left[ \theta(z)~(1-z)~
  +  \theta(-z)~(1+z)  \right]
-{g^2 \over 2(4\pi)^2} ~
 \int_{-1}^0 {d\zeta\over
\zeta}~
\int_{-1}^1 d\zeta'~
{1\over{\zeta'}^{2} m^2 }
~h^{(+)}_{LW}(\zeta,\zeta')
 \times \nonu ~\int_{-1}^0 dz'~\left \{
~\theta (\zeta'-z)~
\theta[z'-\zeta~{ (1+z)\over (1+\zeta')} ]
+
~\theta(z-\zeta')~
\theta[z'-\zeta~{ (1-z)\over (1-\zeta')}  ]
 \right\}=\nonu
 =g^2 ~
\left[ \theta(z)~(1-z)~
  +  \theta(-z)~(1+z)  \right]
+{g^2 \over 2(4\pi)^2} ~\times \nonu
\int_{-1}^1 d\zeta'~
{1\over{\zeta'}^{2} m^2 }
 ~ \left \{
~\theta (\zeta'-z)~{(1+z)\over (1+\zeta')}
+
~\theta(z-\zeta')~{(1-z)\over (1-\zeta')}
 \right\}~a^{(+)}_{LW}(\zeta')
  \label{gWick6} \ee
  where $0\leq (1\pm z)/(1\pm\zeta')<1$ and
  \be
  a^{(+)}_{LW}(z)=\int _{-1}^0 dz'~h^{(+)}_{LW}(z',z)
  \nonu
  a^{(+)}_{LW}(\zeta')=\int _{-1}^0 d\zeta~h^{(+)}_{LW}(\zeta,\zeta')
  \ee
  The S-wave Nakanishi amplitude  is even for $z\to -z$ transformation. 
 
  It should be pointed out that the
  inhomogeneous term has the same form, 
  when the zero-energy limit is performed first, and the
   limit $\mu^2 \to 0$, as briefly shown below.
As a matter of fact, from the inhomogeneous term in Eq. (\ref{gzero2})
(corresponding to $\kappa^2\to 0$), one gets
 \be
\lim_{\mu^2\to 0} {g^2\over \mu^2}  ~\theta(\gamma')
\left [ \theta(z)~\theta(1-z-\gamma'/\mu^2)
 +  \theta(-z)~\theta(1+z-\gamma'/\mu^2)  \right] =\nonu=
{g^2} \delta(\gamma')\left[\theta(z)(1-z) +\theta(-z)(1+z)\right]
\label{WiCue0}\ee
where the the following representation of the delta function has been used
\be
\lim_{\mu^2\to 0}{1\over \mu^2}
\theta(\gamma')~\theta\left [\mu^2(1-|z|)-\gamma'\right]= (1-|z|)~\delta(\gamma')
\label{d16}\ee
Such a relation can be deduced by i) introducing the infinitesimal quantity
$(1-|z|)\mu^2$,
 ii) folding the previous limit with a generic
function $f(\gamma')$ and iii) integrating over $\gamma'$.

  As to the kernel, as above anticipated, it coincides with the  one given
in  Ref. \cite{dae}, for a bound S-wave state, but
  in the limit of zero binding-energy. This comparison, even if it does not apply to a
 physically meaningful case, due to the presence of an essential singularity,
  ${\zeta'}^{2} m^2$,
gives us confidence in our formalism, since it is able
to reproduce the result corresponding to a well known case, after performing 
the non trivial
limits $\mu^2\to 0$ and  then $\kappa^2\to 0$.

Finally, it is important to remark that even the non relativistic 
 Coulomb problem,
for positive energies,
studied through  the Lippman-Schwinger  equation, has an
infrared divergence that must  be treated carefully. In the
Wick-Cutkosky model this problem is present and has to be 
addressed to provide meaningful solutions for the relativistic case
in the scattering region.

\subsection{The S-wave bound state in ladder approximation revisited}
\label{vbrevis}
 The integral equation that
determines the Nakanishi amplitude for a S-wave  bound state, in ladder approximation,
can be rewritten in a simplified form, with respect to the one presented in Ref.
\cite{carbonell1}, once  the uniqueness of the solutions is exploited.
In particular, one can start with  the expression 
of $V_s^{(L)}$ given in Eq. (\ref{vlsbody}), carefully performing the limit
$\zeta \to 0$ and taking a negative value for $\kappa^2$ in the final result.

From Ref. \cite{carbonell1}, one has for a S-wave bound state
\be
\int_0^{\infty}d\gamma'~\frac{g_b(\gamma',z;\kappa^2)}{[\gamma'+\gamma
+z^2 m^2+(1-z^2)\kappa^2-i\epsilon]^2} =\nonu=
\int_0^{\infty}d\gamma''\int_{-1}^{1}d\zeta'\;V_b(\gamma,z;\gamma'',\zeta')
g_b(\gamma'',\zeta';\kappa^2)
\label{newbound}\ee
A direct comparison between Eq. (\ref{vlbound}) and   Eq. (\ref{vlscat}), modulo the value of
$\kappa^2$, leads to the following relation for the interaction kernel
$V^{(L)}_{b}$
\be
V^{(L)}_{b}(\gamma,z;\gamma'',\zeta')= \lim_{\zeta \to 0}
 V^{(L)}_{s}(\gamma,z;\gamma_i,z_i,\gamma'',\zeta,\zeta',cos\theta)= ~- 
~{g^2 \over 2(4\pi)^2}~\times \nonu \lim_{\zeta \to 0}\int^\infty_{-\infty}
 d\gamma' \int_{-1}^1 {dz'}{1 \over  \left[\gamma +z^2 m^2+
\kappa^2(1-z^2)+\gamma'+z'\left({M^2\over 2} z z_i + 2  cos\theta \sqrt{\gamma
\gamma_i}\right)
-i\epsilon   \right]^2}
 \times \nonu \left [{(1+z)\over (1+\zeta'-z_i\zeta)}
~\theta (\zeta'-z-z_i\zeta)~
{\cal Q}'(z,z_i;\gamma'',\gamma',z',\zeta,\zeta',\mu^2)
+\right. \nonu \left . +{(1-z)\over (1-\zeta'+z_i\zeta)}
~\theta
(z-\zeta'+z_i\zeta)~
{\cal Q}'(-z,-z_i;\gamma'',\gamma',z',\zeta,-\zeta',\mu^2)\right]
\ee
where ${\cal Q}'$ can be found in Eq. (\ref{qprimeap}).

Indeed the limit on $\zeta$ must be carefully performed, since  ${\cal Q}'$
 contains two terms, in the first one there is $\theta(\zeta)$ and  
in the second one there is $\theta(-\zeta)$. 
Therefore the limit in the two contributions must be
$\zeta \to 0_+$ and $\zeta \to 0_-$, respectively. 
In what follows this is understood. The relevant limit to be investigated is
(the one corresponding to ${\cal Q}'(-z,-z_i;\gamma'',\gamma',z',\zeta,-\zeta',\mu^2)$ can be carried out analogously)
\be
{\cal H}_b(z;\gamma'',\gamma',\zeta',\mu^2,\kappa^2)=
\lim_{\zeta \to 0} ~{(1+z)\over (1+\zeta'-z_i\zeta)}
~\theta (\zeta'-z-z_i\zeta)~ \times \nonu
\int_{-1}^1 {dz'}{{\cal Q}'(z,z_i;\gamma'',\gamma',z',\zeta,\zeta',\mu^2) \over  \left[\gamma +z^2 m^2+
\kappa^2(1-z^2)+\gamma'+z'\left({M^2\over 2} z z_i + 2  cos\theta \sqrt{\gamma
\gamma_i}\right)
-i\epsilon   \right]^2}
=
\nonu=
\lim_{\zeta \to 0}~{(1+z)\over (1+\zeta'-z_i\zeta)}
~\theta (\zeta'-z-z_i\zeta)~ \times \nonu
\int_{-1}^1 {dz'\over z'}
{\Lambda\left(z,{z'\over \zeta},\zeta,\zeta',\gamma'',\gamma',z_i,\gamma_i,\mu^2\right) \over  \left[\gamma +z^2 m^2+
\kappa^2(1-z^2)+\gamma''+z'\left({M^2\over 2} z z_i + 2  cos\theta \sqrt{\gamma
\gamma_i}\right)
-i\epsilon   \right]^2}
~\times \nonu\theta\left( {1+z\over 1+\zeta'-z_i \zeta}-{z'\over
\zeta}\right)~
\left [
\theta(z')~\theta(\zeta) - \theta(-z')~\theta(-\zeta)
 \right]
=\nonu=
\lim_{\zeta \to 0}{(1+z)\over (1+\zeta'-z_i\zeta)}~
\theta (\zeta'-z-z_i\zeta)~\theta(\zeta)~\int_{0}^{1\over \zeta} {dx\over x}~ 
\theta\left({1+z\over 1+\zeta'-z_i \zeta}-x\right )~\times \nonu
{\Lambda\left(z,x,\zeta,\zeta',\gamma'',\gamma',z_i,\gamma_i,\mu^2\right) \over  \left[\gamma +z^2 m^2+
\kappa^2(1-z^2)+\gamma'+x \zeta\left({M^2\over 2} z z_i + 2  cos\theta \sqrt{\gamma
\gamma_i}\right)
-i\epsilon   \right]^2}
+
\nonu -
\lim_{\zeta \to 0}  {(1+z)\over (1+\zeta'-z_i\zeta)}~
\theta (\zeta'-z-z_i\zeta)~ \theta(-\zeta)~\int^{0}_{-{1\over \zeta}} {dx\over
x}~\theta\left({1+z\over 1+\zeta'-z_i \zeta}-x\right )~\times \nonu 
 {\Lambda\left(z,x,\zeta,\zeta',\gamma'',\gamma',z_i,\gamma_i,\mu^2\right) \over  \left[\gamma +z^2 m^2+
\kappa^2(1-z^2)+\gamma'+x \zeta\left({M^2\over 2} z z_i + 2  cos\theta \sqrt{\gamma
\gamma_i}\right)
-i\epsilon   \right]^2}   
\ee
where the change of variable $x=z'/ \zeta$ has been inserted. By recalling that
\be
{1\over|\zeta|}\geq {1+z\over 1+\zeta'-z_i\zeta}
\ee
since $1\geq |\zeta|$ and $\theta(\zeta'-z_i\zeta-z)$ one can write
\be
{\cal H}_b(z;\gamma'',\gamma',\zeta',\mu^2,\kappa^2)
=
\lim_{\zeta \to 0}{(1+z)\over (1+\zeta'-z_i\zeta)}~\theta (\zeta'-z-z_i\zeta)~
\left[\theta(\zeta)+\theta(-\zeta)\right]~\times \nonu
  \int_{0}^{1+z\over 1+\zeta'-z_i\zeta} {dx\over x}
  {\Lambda\left(z,x,\zeta,\zeta',\gamma'',\gamma',z_i,\gamma_i,\mu^2\right) 
  \over  \left[\gamma +z^2 m^2+
\kappa^2(1-z^2)+\gamma'+x \zeta\left({M^2\over 2} z z_i + 2  cos\theta \sqrt{\gamma
\gamma_i}\right)
-i\epsilon   \right]^2}
\ee
where the following identity has been exploited
\be
-\theta(-\zeta) \int^{0}_{-{1\over \zeta}} {dx\over x}=
\theta(-\zeta) \int_{0}^{1\over |\zeta|} {dx\over x}
\ee
Then, one gets 
\be 
{\cal H}_b(z;\gamma'',\gamma',\zeta',\mu^2,\kappa^2)={(1+z)\over (1+\zeta')}~
~{\theta (\zeta'-z) \over  \left[\gamma +z^2 m^2+
\kappa^2(1-z^2)+\gamma'
-i\epsilon   \right]^2}
 \times \nonu   
\int_{0}^{(1+z)\over (1+\zeta')} {dx\over x}~
\tilde\Lambda\left(z,x,\zeta',\gamma'',\gamma',\mu^2,\kappa^2\right)
\ee
where
\be
\tilde\Lambda\left(z,x,\zeta',\gamma'',\gamma',\mu^2,\kappa^2\right)=
\sum_{i=\pm} {\partial \over \partial \lambda} \tilde y_i(0)
~{1 \over|\tilde y^2_i(0){\cal A}_b(\zeta',\gamma'',\kappa^2)
 -\mu^2|} ~ \times \nonu\left[ \delta(\tilde y_i(0)) y^2_i(0)
  -\theta(\tilde y_i(0)) ~ { 2\mu^2~ \tilde y_i(0) \over\tilde  y^2_i(0){\cal
  A}_b(\zeta',\gamma'',\kappa^2)
 -\mu^2}\right]
\ee
with
\be
\tilde y_\pm(0)= 
{1 \over 2{\cal A}_b(\zeta',\gamma'',\kappa^2)} ~\times \nonu \left[
 -{\cal B}_b(z,x,\zeta',\gamma'',\gamma',\mu^2,0) 
 \pm \sqrt{{\cal B}_b^2(z,z',z_i,\zeta,\zeta',\gamma'',\gamma',\mu^2,0) 
- 4 \mu^2 {\cal
A}_b(\zeta',\gamma'',\kappa^2)} \right]
\nonu
{\partial \over \partial \lambda}\tilde y_i(0)=~\mp~ 
{(1+\zeta')\over (1+z)}~~ 
{
~ \tilde y_\pm(0)\over 
\sqrt{{\cal B}_b^2(z,x,\zeta',\gamma'',\gamma',\mu^2,0) 
- 4 \mu^2 {\cal A}_b(\zeta',\gamma'',\kappa^2)}}
\ee
and
\be
{\cal A}_b(\zeta',\gamma'',\kappa^2)={\zeta'}^{2} \frac{M^2}{4} + 
\kappa^2 +\gamma''
\nonu
{\cal B}_b(z,x,\zeta',\gamma'',\gamma',\mu^2,0)
=\mu^2 +\gamma''-{\gamma' \over x}
\ee
 The positivity of $\gamma''$ in Eq. (\ref{newbound}) leads to the positivity of ${\cal A}_b$.
This implies that ${\cal B}_b$ be negative (cf the end of Appendix \ref{vlad}), and eventually $\gamma'$ be
positive, consistently with Eq. (\ref{newbound}), analyzed in Ref. 
\cite{carbonell1}.  This could also be expected on physical grounds, since the particle
production, or a cut in the mathematical language, must be avoided in the bound
state kernel $V^{(L)}_{b}(\gamma,z;\gamma'',\zeta')$, namely in the denominator
appearing in Eq. (\ref{newvb}).
Therefore the interaction kernel $V^{(L)}_b$ becomes
\be
V^{(L)}_{b}(\gamma,z;\gamma'',\zeta')=-{g^2 \over 2(4\pi)^2}~
 \int^\infty_{0}
 d\gamma' {1 \over  \left[\gamma +z^2 m^2+
\kappa^2(1-z^2)+\gamma'
-i\epsilon   \right]^2}
 \times \nonu ~\left\{{(1+z)\over (1+\zeta')}
~\theta (\zeta'-z)
\int_{0}^{1+z\over 1+\zeta'} {dx\over x}~   
\tilde\Lambda\left(z,x,\zeta',\gamma'',\gamma',\mu^2,\kappa^2\right)
+\right. \nonu \left. +
{(1-z)\over (1-\zeta')}
~\theta (z-\zeta')
\int_{0}^{1-z\over 1-\zeta'} {dx\over x}~  
\tilde\Lambda\left(-z,x,-\zeta',\gamma'',\gamma',\mu^2,\kappa^2\right)
\right\}
\label{newvb}\ee
The new form of the integral integration for $g_b(\gamma',z;\kappa^2)$ is given by
\be
g_b(\gamma',z;\kappa^2)=-{g^2 \over 2(4\pi)^2}~
\int_0^{\infty}d\gamma''\int_{-1}^{1}d\zeta'~
g_b(\gamma'',\zeta';\kappa^2)~\times \nonu
 \times \nonu ~\left\{{(1+z)\over (1+\zeta')}
~\theta (\zeta'-z)
\int_{0}^{1+z\over 1+\zeta'} {dx\over x}~   
\tilde\Lambda\left(z,x,\zeta',\gamma'',\gamma',\mu^2,\kappa^2\right)
+\right. \nonu \left. +
{(1-z)\over (1-\zeta')}
~\theta (z-\zeta')
\int_{0}^{1-z\over 1-\zeta'} {dx\over x}~  
\tilde\Lambda\left(-z,x,-\zeta',\gamma'',\gamma',\mu^2,\kappa^2\right)
\right\}
\ee

\section{Conclusion}\label{concl}

 We analyzed the Bethe-Salpeter equation for scattering states by
exploiting  the Nakanishi perturbation theory integral representation of the
multi-leg transition amplitudes\cite{nak63}. In this way, one has
the possibility to
single out their analytic behavior in Minkowski space
 and    to exactly
 project (namely explicitly integrating over the variable $k^-$) any
  multi-leg amplitude onto the null plane, so that  the relative
 Light-front time (between free legs) can be eliminated.    In this context, we have addressed
 the  BSE for
  the scattering states, in  Minkowski space, extending the work of Ref.
  \cite{carbonell1,carbonell2}, where
 the  explicitly-covariant Light-front approach was applied to
obtain
 bound states for a massive two-scalar system, interacting through a massive
 scalar exchange.

  A key ingredient of our work is the one-to-one correspondence
   between the BS amplitude and the LF valence wave
  function
  of the interacting system. Such a relation is implemented  
  through an operator that
  is able to produce the full complexity of the Fock space on top 
  of the valence
  component \cite{sales00,sales01,hierareq,adnei07,adnei08,Tobias_FB}, and
  noteworthily it makes feasible the study of the Nakanishi
  amplitude  through the valence component, without loosing
  any physical content.
  
  The LF projection technique allows us to write down an exact 3D equation for the valence component from the 4D
  BSE, both for bound and scattering states. In turn, given the relation between
  BS amplitude and  valence component on one side, and  the Nakanishi weight
  function on the other side, one can determine the last one, through a more
  simple mathematical treatment, namely by using the 3D equation for the
  valence component without facing with the complexity of the  4D
  Minkowski space (see for the bound case \cite{KW}).

In the case of  scattering states, it has been necessary to consider the PTIR of
the half-off-shell T-matrix, i.e. a four-leg amplitude (while, for the bound
state,  it is needed the PTIR of the vertex function
 \cite{carbonell1,carbonell2}). Then, applying the projection method,
 the four-dimensional
 inhomogeneous BSE has been exactly reduced to a three-dimensional equation for
 the valence component, that in turn  it has allowed  to devise
 an  equation for
the Nakanishi weight
 function (a real function),  without angular momentum decomposition.  
 The relevant equation, for a massive 
 two-scalar system, interacting through a massive
 scalar exchange in the continuum, is Eq. (\ref{bsnewscatt}), that for the sake of clarity
 we report in this Conclusion, viz
 \be
\int_{-1}^1 dz'\int_{-\infty}^{\infty}d\gamma'
\frac{g^{(+)}(\gamma',z',z;\gamma_i,z_i)}
{[\gamma'+\gamma+ z^{ 2}
m^2+(1- z^{ 2})\kappa^2
 +{M \over 2} z~z' ({M \over 2} z_i + k^-_i ) +
 2 z'cos \theta \sqrt{\gamma
\gamma_i}-i\epsilon]^2}=\nonu= {\cal
I}^{LF}(\gamma,z;\gamma_i,z_i,cos \theta) +\nonu +
\int_{-\infty}^{\infty}d\gamma'\int_{-1}^{1}d\zeta\int_{-1}^{1}d\zeta'
~ V^{LF}_s(\gamma,z;\gamma_i,z_i,\gamma',\zeta,\zeta',cos \theta)
~g^{(+)}(\gamma',\zeta,\zeta';\gamma_i,z_i) \nonumber \ee
where the inhomogeneous term, ${\cal I}^{LF}$,  and the kernel, $V^{LF}_s$, are
given in Eqs. (\ref{inlf}) and (\ref{VS}), respectively.
As a by-product, we have obtained the scattering amplitude in terms of  the
$g^{(+)}$,  ${\cal I}^{LF}$  and  $V^{LF}_s$. This relation will be useful for
phenomenological studies.

 The explicit expression of the previous integral equation  has been 
 also obtained for the 4D BS kernel
 in ladder approximation, preparing the matter for
 forthcoming numerical investigations. In particular, a simpler integral equation for
 determining the Nakanishi amplitude, in ladder approximation, has been worked
 out by explicitly applying uniqueness, 
  see  Eq. (\ref{gladder}).  It is
worth noting that, in ladder approximation,  the validity of the
uniqueness in the continuum can be   checked by obtaining the
Nakanishi weight function, through the two possibilities given by
the integral equation in (\ref{ladscatc}) and the one in
(\ref{gladder}). Moreover, it should be emphasized, from one side, the
simplicity and the benefit of the LF method for obtaining the kernel
of the integral equation for  scattering states, and from the other side 
the elaborated mathematical steps necessary to  explicitly use
 uniqueness in order to simplify the form of integral equation (\ref{ladscatc})
 (see e.g. the work of ref.
\cite{KW}).

We explored our formalism in ladder approximation, by investigating
two limits: i) the zero energy limit $\kappa^2\to 0$ and ii) the
Wick-Cutkosky model in the continuum, that corresponds to  an exchanged
massless  scalar, namely $\mu^2\to 0$, obtaining
for this case a separable form for the Nakanishi weight function. 
The cross-ladder contribution will be
considered elsewhere.  Moreover, a new form of the ladder kernel for bound states
has been provided, allowing a simplified form for the integral equation
determining the Nakanishi weight.

Nakanishi conjectured \cite{nak642} that his approach could be valid beyond the
perturbative regime. Indeed, numerical solutions of the bound-state 
problem have 
shown that, for truncated  kernels, his method can be successfully applied. The
investigation of the validity of the Nakanishi approach to
non perturbative problems in the continuum, as  a reasonable
extension of what has been already  done for the bound-state poles of
the transition matrix, can be performed by solving   the
integral equations obtained in this work.

Finally, the Nakanishi PTIR, applied to bound and scattering states, opens 
the possibilities of studying many new issues, since it
 is not constrained to  3+1 dimensions or  to two interacting particles.
 For instance, let us mention  that
 it could be applied to
three-interacting bosons (see  e.g. \cite{patricia} for the three-boson BSE 
and the investigation of the final state
interaction in  three-body decays of heavy mesons), 
or to consider 
the extension to 2+1 dimensions  and fermionic systems (useful, e.g., to treat 
the Dirac electrons in graphene, for reviews see
\cite{CastroNeto:2009zz,Peres2010}).

\section*{Acknowledgments}
TF and MV  acknowledge the hospitality of INFN Sezione di Roma. TF also thanks 
 the partial financial 
support from 
 the Conselho Nacional
de Desenvolvimento Cient\'{\i}fico e Tecnol\'{\i}ogico (CNPq),
the  Funda\c c\~ao de Amparo \`a Pesquisa do
Estado de S\~ao Paulo (FAPESP) and 
the Italian MUR through the PRIN 2008.
\appendix
\section{The inhomogeneous term in ladder approximation}\label{inho}
The inhomogeneous term,  present in the 3D LF equation for scattering
states (cf Eq. (\ref{bsnewscatt})), is given by
\be
{\cal I}^{LF}(\gamma,z;\gamma_i,z_i,cos \theta)= p^+~~\int {dk^-\over 2 \pi}G_0^{(12)}(k,p)
~i{\cal K}(k,k_i,p)=\nonu=~i^2~p^+~\int {dk^-\over 2 \pi}
{1 \over \left[(\frac{p}{2}+ k)^2-m^2+i\epsilon\right]}~
{1 \over\left[(\frac{p}{2}-k)^2-m^2+i\epsilon\right]}
~i{\cal K}(k,k_i,p)
\ee
Let us calculate the term in ladder approximation.
Within such an approximation, the kernel $i{\cal K}$ is given by (cf Ref.
\cite{carbonell1})
\be
i{\cal K}^{(L)} (k,k_i)={i (-ig)^2 \over (k-k_i)^2 -\mu^2 +i\epsilon}
\ee
where $\mu$ is the mass of the exchanged scalar ($g^2=16\pi m^2 \alpha$) and
the momentum transfer is given by
$p_{1}-p_{1i}=(p/2)+k-(p/2)-k_i$.
Let us recall that $k_i=(p_{1i} -p_{2i})/2$ and
$p=p_{1i} +p_{2i}=p_{1} +p_{2}$.

Then, in a reference frame where $\mbf{p}_\perp=0$ and $p^{\pm}=M$, one has
\be{\cal I}^{(L)}(\gamma,z;\gamma_i,z_i,cos \theta)=
i g^2 ~p^+~
\times \nonu \int {dk^-\over 2 \pi}
{1 \over \left[(\frac{p}{2}+ k)^2-m^2+i\epsilon\right]}~
{1 \over\left[(\frac{p}{2}-k)^2-m^2+i\epsilon\right]}~
{1 \over (k-k_i)^2 -\mu^2 +i\epsilon}=\nonu=
i g^2 ~p^+~
 \int {dk^-\over 2 \pi}~{1 \over (M/2+k^+)~(M/2-k^+)~(k^+-k^+_i)}~\times
\nonu{1 \over \left[(\frac{p}{2}+ k)^- - (\frac{p}{2}+ k)^-_{on}+i\epsilon/(M/2+k^+)\right]}~
{1 \over\left[(\frac{p}{2}- k)^- - (\frac{p}{2}- k)^-_{on}+i\epsilon/(M/2-k^+)\right]}~
\times \nonu{1 \over (k-k_i)^- -(k-k_i)^-_{on} +i\epsilon/(k^+-k^+_i)}
=\nonu =
-i g^2 ~p^+~
 \int {dk^-\over 2 \pi}~{8 \over M^3(1-z^2)~(z-z_i)}~\times
\nonu{1 \over \frac{M}{2}+ k^- - (\frac{p}{2}+ k)^-_{on}+i2\epsilon/[M/(1-z)]}~~
{1 \over\frac{M}{2}- k^- - (\frac{p}{2}- k)^-_{on}+i2\epsilon/[M (1+z)]}~
\times \nonu{1 \over k^- -k_i^- -(k-k_i)^-_{on} -i2\epsilon/[M(z-z_i)]}
\label{intinh}\ee
with $z=-2k^+/M$, $z_i=-2k_i^+/M=2k_i^-/M$ (since $p\cdot k_i=0$) and
\be
(\frac{p}{2}+ k)^-_{on}=p^-_{1on}={2(m^2+\gamma) \over M (1-z)}
\nonu
(\frac{p}{2}- k)^-_{on}=p^-_{2on}={2(m^2+\gamma) \over M (1+z)}
\nonu
(k-k_i)^-_{on}=-{2(\mu^2+\gamma+\gamma_i -2cos\theta~\sqrt{\gamma \gamma_i)} \over M(z-z_i)}
\ee
where $cos \theta =\widehat{\bf k}_\perp \cdot \widehat{\bf k}_{i\perp}$, since we
have chosen $\widehat x=\widehat{\bf k}_{i\perp}$,   $\gamma=k^2_{\perp} $ and $\gamma_i=k^2_{i\perp}$.
It should be pointed out that $1\ge |z|$ and
$1 \ge |z_i|$ since we are considering only particles, i.e $(p^+/2) \pm k^+\ge 0$
and $(p^+/2) \pm k^+_i\ge 0$.

In order to perform the analytic integration, one has to consider the
following three poles
\be
k^-_{L}={2(m^2+ \gamma) \over {M}(1-z)}-{M\over 2}-i{2 \epsilon\over  M(1-z)}
\nonu
k^-_{U}=-{2(m^2+ \gamma)\over  {M}(1+z)}+{M\over 2}+i
{2 \epsilon\over  M(1+z)} \nonu
k^-_{LU}=k^-_i- {2(\gamma+ \gamma_i+\mu^2   -
2cos\theta~\sqrt{\gamma \gamma_i})\over M(z-z_i)}  +i{2\epsilon \over M(z-z_i)}=\nonu
={M\over 2} z_i- {2(\gamma+ \gamma_i+\mu^2   -
2cos\theta~\sqrt{\gamma \gamma_i})\over M(z-z_i)}  +i{2\epsilon \over M(z-z_i)}
\ee

First of all, let us mention that the cases $|z|=1$ and $|z_i|=1$ lead to a vanishing contributions,
as one can check by properly recombining the factors
$1/(1-z^2)$ and $1/(z-z_i)$ in the integral (\ref{intinh}) and the previous poles. When $1>z>z_i>-1$ one can close the integration
 contour in the
lower plane, taking the residue at $k^-=k^-_L$. While for
$ 1>z_i>z>-1$ one can choose  $k^-=k^-_U$.
Then, one has
 \be {\cal I}^{(L)}(\gamma,z;\gamma_i,z_i,cos \theta)=i g^2 ~
 \int {dk^-\over 2 \pi}~{8 \over M^2(1-z^2)~(z-z_i)}~\times
\nonu{1 \over \left( k^- -k^-_L\right)}~
{1 \over\left( k^- - k^-_U\right)}~
{1 \over \left( k^- -k_{LU}\right)}
 =\nonu=i^2 g^2 ~
~{8\over {M^2} (1 -z^2)~(z-z_i)}~
\left\{-
~{\theta(z-z_i) \over {2(m^2+\gamma)\over M(1-z)}
-{M\over 2}+{2(m^2+\gamma)\over M(1+z)}
-{M\over 2}-i\epsilon}~\times \right.
\nonu \left. {1 \over  {2(m^2+\gamma)\over M(1-z)}
-{M\over 2}(1+z_i)   + {2(\gamma+ \gamma_i+\mu^2   -
2cos\theta~\sqrt{\gamma \gamma_i})\over M(z-z_i)}  -i\epsilon}
+\right.
\nonu \left.+
~{\theta(z_i-z) \over -{2(m^2+\gamma)\over M(1+z)}
+{M\over 2}-{2(m^2+\gamma)\over M(1-z)}
+{M\over 2}+i\epsilon}~\times \right.
\nonu \left. {1 \over  -{2(m^2+\gamma)\over M(1+z)}
+{M\over 2}(1-z_i)   + {2(\gamma+ \gamma_i+\mu^2   -
2cos\theta~\sqrt{\gamma \gamma_i})\over M(z-z_i)}  +i\epsilon}
\right\}\ee
Finally,
 one can rewrite\be
{\cal I}^{(L)}(\gamma,z;\gamma_i,z_i,cos \theta)=-g^2 ~{8\over {M^2} (1 -z^2)~(z-z_i)}~
{1 \over M-{2(m^2+\gamma)\over M(1+z)}
-{2(m^2+\gamma)\over M(1-z)}
+i\epsilon}\times \nonu
\left\{
{\theta(z-z_i) \over  {2(m^2+\gamma)\over M(1-z)}
  -{M\over 2} (1+z_i) + {2(\gamma+ \gamma_i+\mu^2   -
2cos\theta~\sqrt{\gamma \gamma_i})\over M(z-z_i)}  -i\epsilon}
+\right.
\nonu \left.+
 { \theta(z_i-z)\over  -{2(m^2+\gamma)\over M(1+z)}+{M\over2} (1-z_i) +
  {2(\gamma+ \gamma_i+\mu^2   -
2cos\theta~\sqrt{\gamma \gamma_i})\over M(z-z_i)}  +i\epsilon}
\right\}
=\nonu=
g^2 ~
{1 \over  \gamma +(1-z^2)\kappa^2 +z^2 m^2
-i\epsilon}~{1\over  (z-z_i)}~\times \nonu
\left\{
{\theta(z-z_i) \over  {(m^2+\gamma)\over (1-z)}
  -{M^2\over 4} (1+z_i) + {(\gamma+ \gamma_i+\mu^2   -
2cos\theta~\sqrt{\gamma \gamma_i})\over (z-z_i)}  -i\epsilon}
+\right.
\nonu \left.+
 { \theta(z_i-z)\over  -{(m^2+\gamma)\over (1+z)}+{M^2\over 4} (1-z_i) +
  {(\gamma+ \gamma_i+\mu^2   -
2cos\theta~\sqrt{\gamma \gamma_i})\over (z-z_i)}  +i\epsilon}
\right\}=
\nonu=~g^2 ~
{1 \over   \gamma +(1-z^2)\kappa^2 +z^2 m^2-i\epsilon}~
{\cal G}^{(L)}(\gamma,z;\gamma_i,z_i,cos \theta)\ee
where ${\cal G}^{(L)}(\gamma,z;\gamma_i,z_i,cos \theta)$ is given by
\be
{\cal G}^{(L)}(\gamma,z;\gamma_i,z_i,cos \theta)=~
{1\over  (z-z_i)}~
\left\{
{\theta(z-z_i) \over  {(m^2+\gamma)\over (1-z)}
  -{M^2\over 4} (1+z_i) + {(\gamma+ \gamma_i+\mu^2   -
2cos\theta~\sqrt{\gamma \gamma_i})\over (z-z_i)}  -i\epsilon}
+\right.
\nonu \left.+
 { \theta(z_i-z)\over  -{(m^2+\gamma)\over (1+z)}+{M^2\over 4} (1-z_i) +
  {(\gamma+ \gamma_i+\mu^2   -
2cos\theta~\sqrt{\gamma \gamma_i})\over (z-z_i)}  +i\epsilon}
\right\}=
\nonu
 ={\theta(z-z_i) (1-z)\over  (z-z_i)\left[m^2+\gamma
  -{M^2\over 4} (1+z_i)(1-z)\right] +
 (1-z) (\gamma+ \gamma_i+\mu^2  - 2cos\theta~\sqrt{\gamma \gamma_i})  -i\epsilon}
+
\nonu +
 { \theta(z_i-z)(1+z)\over  (z-z_i)
 \left[{M^2\over 4} (1-z_i)(1+z)-(m^2+\gamma)\right] +
 (1+z)(\gamma+ \gamma_i+\mu^2 -2cos\theta~\sqrt{\gamma \gamma_i})  -i\epsilon}
 =
 \nonu=
{\theta(z-z_i) (1-z)\over  \beta(z,z_i)+\gamma(1-z_i) - 2(1-z)cos\theta~\sqrt{\gamma
\gamma_i}
  -i\epsilon}
+
\nonu +
 { \theta(z_i-z)(1+z)\over  \beta(-z,-z_i)+\gamma (1+z_i)-
2(1+z)cos\theta~\sqrt{\gamma \gamma_i}
   -i\epsilon}
\ee
where  $\beta(z,z_i)$is given by:
\be
\beta(z,z_i)=(1-z)\left[\gamma_i+\mu^2-(z-z_i)(1+z_i) {M^2\over
4}-m^2\right ]+(1-z_i)m^2=
\nonu=
(1-z)\left[\mu^2+{M^2\over
4}(1-z)(1+z_i) -2m^2\right ]+(1-z_i)m^2
\label{cd5}
\ee
where $ M^2/4=(m^2 +\gamma_i)/(1-z^2_i)$ has been used.
 For further purposes, the denominator can written as
\be
\beta(z,z_i)+\gamma(1-z_i) =\nonu=
(1-z_i)[\gamma +(1-z^2) \kappa^2 +z^2 m^2] +
\beta(z,z_i) -(1-z_i) [(1-z^2) \kappa^2 +z^2 m^2]=
\nonu=
 (1-z_i)A +(1-z)\left[\mu^2+{M^2\over
4}(1-z)(1+z_i) -2m^2\right ] +(1-z_i) (1-z^2){M^2\over 4}
=
\nonu=
 (1-z_i)A +(1-z)B \ee
where $A=\gamma +(1-z^2) \kappa^2 +z^2 m^2$ and $B=\mu^2 -2\kappa^2- z
z_i{M^2/2}$

By using the Feynman trick to reduce
${\cal I}^{(L)}(\gamma,z;\gamma_i,z_i,cos \theta)$ to a single
 denominator,
one has
\be
{\cal I}^{(L)}(\gamma,z;\gamma_i,z_i,cos \theta)=
~ g^2 ~ \int_0^1d\alpha
~\times \nonu \left[ {\theta(z-z_i)(1-z) \over
\left[A\alpha +\left(
(1-z_i)A +
(1-z)B - 2(1-z)cos\theta~\sqrt{\gamma
\gamma_i}
\right)(1-\alpha) -i\epsilon\right]^2 }\right. +\nonu
 \left.+ { \theta(z_i-z)(1+z)\over \left[A\alpha +\left((1+z_i)A+
 (1+z)B-
2(1+z)cos\theta~\sqrt{\gamma \gamma_i}\right)(1-\alpha) -i\epsilon
\right]^2} \right\} =\nonu
=
 g^2 ~ \int_0^1{d\alpha\over [1 -z_i (1 -\alpha)]^2}
~  {\theta(z-z_i)(1-z) \over
\left[A +\left(
B - 2cos\theta~\sqrt{\gamma
\gamma_i}
\right){(1-z)(1-\alpha)\over[1 -z_i (1 -\alpha)]}  -i\epsilon\right]^2 } +\nonu +
g^2 ~ \int_0^1{d\alpha\over [1 +z_i (1 -\alpha)]^2}
 { \theta(z_i-z)(1+z)\over \left[A +\left(
 B-
2cos\theta~\sqrt{\gamma \gamma_i}\right){(1+z)(1-\alpha)\over[1 +z_i (1
-\alpha)]}  -i\epsilon
\right]^2}
\ ,
\ee
where
$[1 \pm z_i (1-\alpha)]\geq 0$ since $1>|z_i|$ ($|z_i|=1$ can be excluded for
physical motivation, since this amounts to an infinite energy for one of the
incoming particle). The change of variable in the
first integral is
$$y = {(1-z)(1-\alpha)\over[1 -z_i (1 -\alpha)]} $$ and in the second one is
$$y = {(1+z)(1-\alpha)\over[1 +z_i (1 -\alpha)]} $$
Then, one has
\be{\cal I}^{(L)}(\gamma,z;\gamma_i,z_i,cos \theta)=
 g^2 ~ \int_0^{1} dy
~  {\theta(z-z_i) ~\theta \left[1-z -y(1-z_i)\right]\over
\left[A + yB - 2ycos\theta~\sqrt{\gamma
\gamma_i}  -i\epsilon\right]^2 } +\nonu +
g^2 ~ \int_0^1 dy
 { \theta(z_i-z)~\theta \left[1+z -y(1+z_i)\right]\over \left[A +
 yB-
2ycos\theta~\sqrt{\gamma \gamma_i}  -i\epsilon
\right]^2}=
\nonu=
g^2 ~ \int_{-1}^{1} dy~ \theta(y)
~ \left\{ {\theta(z-z_i) ~\theta \left[1-z -y(1-z_i)\right]\over
\left[A + yB - 2ycos\theta~\sqrt{\gamma
\gamma_i}  -i\epsilon\right]^2 } + \right.\nonu +
\left.
 { \theta(z_i-z)~\theta \left[1+z -y(1+z_i)\right]\over \left[A +
 yB-
2ycos\theta~\sqrt{\gamma \gamma_i}  -i\epsilon
\right]^2}\right\}
\ee

Let us change $y \to -y$, then one gets
\be
 {\cal I}^{(L)}(\gamma,z;\gamma_i,z_i,cos \theta)=
g^2 ~ \int_{-1}^{1} dy~ \theta(-y)
~ \left\{ {\theta(z-z_i) ~\theta \left[1-z +y(1-z_i)\right]\over
\left[A - yB + 2ycos\theta~\sqrt{\gamma
\gamma_i}  -i\epsilon\right]^2 } + \right.\nonu +
\left.
 { \theta(z_i-z)~\theta \left[1+z +y(1+z_i)\right]\over \left[A -
 yB+
2ycos\theta~\sqrt{\gamma \gamma_i}  -i\epsilon
\right]^2}\right\}=
\nonu=
g^2 ~ \int_{-1}^{1} dy~\int^\infty_{-\infty}~d\gamma'~ {\theta(-y)
~ \delta(\gamma' -\gamma_a(y))
\over
\left[\gamma'+\gamma+(1-z^2)\kappa^2 +z^2m^2 +y{M^2 \over 2} z  z_i +2ycos\theta~\sqrt{\gamma
\gamma_i}  -i\epsilon\right]^2 } ~\times \nonu
\left\{ \theta(z-z_i) ~\theta \left[1-z +y(1-z_i)\right]
 + \theta(z_i-z)~\theta \left[1+z +y(1+z_i)\right]\right\}
\ , \label{cd4n}
\ee
where the term $y(M^2 /2) z  z_i$ has been inserted for obtaining the same
denominator in Eq. (\ref{ladscat}) and
\be
\gamma_a(y)=-yB-y{M^2 \over 2} z  z_i =y(2\kappa^2-\mu^2)
\label{cd7}\ee

\section{The kernel $V^{LF}_s$ in ladder approximation}\label{vlad}
In  the ladder approximation of the 3D LF equation for bound states
(cf Eq. (\ref{ptireq})) it is present the following  kernel $V^{(L)}_b$
\be
V^{(L)}_b(\gamma,z;\gamma',z')=
~i^2~g^2p^+\int_{-\infty}^{\infty}{d k^- \over 2\pi}~
{1 \over \left[(\frac{p}{2}+k)^2-m^2+i\epsilon\right]}
~ {1\over \left[(\frac{p}{2}-k)^2-m^2+i\epsilon\right]} \times
\nonu
\int \frac{d^4k''}{(2\pi)^4}\frac{1}
{\left[{k''}^2+p\cdot k'' z' -\gamma'-\kappa^2+i\epsilon\right]^3}
~{1\over (k-k'')^2 -\mu^2 +i\epsilon}=\nonu=
~-~g^2p^+
\int \frac{d^4k''}{(2\pi)^4}\frac{1}
{\left[{k''}^2+p\cdot k'' z'-\gamma'-\kappa^2+i\epsilon\right]^3}
~\times \nonu \int_{-\infty}^{\infty}{d k^- \over 2\pi}~
{1 \over \left[(\frac{p}{2}+k)^2-m^2+i\epsilon\right]}
~ {1\over \left[(\frac{p}{2}-k)^2-m^2+i\epsilon\right]}{1\over (k-k'')^2
-\mu^2 +i\epsilon}
\label{vlbound}\ee
An explicit expression can be found in Ref. \cite{carbonell1}. Another
explicit expression can be obtained from $V^{(L)}_s$ as given in subsect.
\ref{vbrevis}, and recalling that
$
V_b(\gamma,z;\gamma',\zeta')=\lim _{\zeta\to 0}V_s(\gamma,z;\gamma_i,z_i,\gamma',\zeta,\zeta',cos\theta)$.

In this Appendix, it will be illustrated in detail the kernel for the
scattering-state equation (cf Eq. (\ref{bsnewscatt})). In this case,
the kernel acquaints new dependencies: i) upon another compact variable
 $z''$ and ii)
the  $cos \theta= \widehat{\mbf k}_\perp \cdot \widehat{\mbf k}_{i \perp}$.
Moreover, there is
an {\em understood dependence} upon $\gamma_i$, $z_i$ and $p$. In ladder
approximation, one has the
following expression for $V_s^{(L)}$
\be
V^{(L)}_{s}(\gamma,z;\gamma_i,z_i,\gamma',\zeta,\zeta',cos\theta)=~-
p^+~\int_{-\infty}^{\infty}{d k^- \over 2\pi}~G_0^{(12)}(k,p)
~\times \nonu\int \frac{d^4k''}{(2\pi)^4}\frac{{\cal K}^{(L)}(k,k'',p)}
{\left[k^{\prime\prime 2}+{p^2\over 4}-m^2 +p\cdot k''~\zeta^\prime
 +2 k''\cdot k_i~\zeta -\gamma' +i\epsilon \right ]^3}=\nonu=
~-~g^2p^+\int_{-\infty}^{\infty}{d k^- \over 2\pi}~
{1 \over \left[(\frac{p}{2}+k)^2-m^2+i\epsilon\right]}
~ {1\over \left[(\frac{p}{2}-k)^2-m^2+i\epsilon\right]} \times
\nonu
\int \frac{d^4k''}{(2\pi)^4}\frac{1}
{\left[{k''}^2+p\cdot k'' \zeta'+2 k''\cdot k_i \zeta -\gamma'-\kappa^2+i\epsilon\right]^3}
~{1\over (k-k'')^2 -\mu^2 +i\epsilon}=\nonu
=~-~g^2p^+\int_{-\infty}^{\infty}{d k^- \over 2\pi}~
{1 \over \left[(\frac{p}{2}+k)^2-m^2+i\epsilon\right]}
~ {1\over \left[(\frac{p}{2}-k)^2-m^2+i\epsilon\right]}~\times \nonu
{\cal P}(k,\gamma',\zeta,\zeta',cos \theta)
\label{vlscat}\ee
where
\be
{\cal P}(k,\gamma',\zeta,\zeta',cos \theta)=
\int \frac{d^4k''}{(2\pi)^4}\frac{1}
{\left[{k''}^2+p\cdot k'' \zeta'+2 k''\cdot k_i \zeta -\gamma'-\kappa^2+i\epsilon\right]^3}
~\times \nonu{1\over (k-k'')^2 -\mu^2 +i\epsilon}
\ee
 Adopting the formula (see also Ref. \cite{carbonell1})
 \be \frac{1}{a
b^3}=\int_0^1 \frac{3v^2 d v}{[a(1-v)+b v]^4}
\ee
where
\be
a=(k-k'')^2 -\mu^2 +i\epsilon
\nonu
b={k''}^2+p\cdot k'' \zeta'+2 k''\cdot k_i \zeta -\gamma'-\kappa^2+i\epsilon
\nonu
(1-v)a+vb=(1-v)[(k-k'')^2 -\mu^2]+v[{k''}^2+p\cdot k'' \zeta'+
2 k''\cdot k_i \zeta
-\gamma'-\kappa^2]=\nonu=
{k''}^2- 2k'' \cdot [(1-v)k-v\zeta'{p \over 2} -v k_i \zeta]
+(1-v)(k^2 -\mu^2) -v(\gamma'+\kappa^2)+i\epsilon=\nonu=
q^2-[(1-v)k-v\zeta' \frac{p}{2} -v k_i \zeta]^2+(1-v)(k^2 -\mu^2) -v(\gamma'+\kappa^2)+i\epsilon\ee
with
 $$ q=k''-(1-v)k+v\zeta' \frac{p}{2} +v k_i \zeta$$
Then one has \be {\cal P}(k,\gamma',\zeta,\zeta',cos \theta)=
{1\over (2\pi)^4}\int_0^1 3v^2 d v
\int\frac{d^4q}{\left[{q}^2+A(k,\zeta,\zeta',cos
\theta)+i\epsilon\right]^4} \nonu ={i \over 2(4\pi)^2}\int_0^1
\frac{v^2 d v}{[A(k,\zeta,\zeta',cos \theta)+i\epsilon]^2} \ee where
\be A(k,\zeta,\zeta',cos \theta)=-[(1-v)k-v\zeta' \frac{p}{2} -v k_i
\zeta]^2+(1-v) (k^2 -\mu^2) -v(\gamma'+\kappa^2) =\nonu=k^2 v(1-v)+2
(1-v) v ~k \cdot [\zeta' \frac{p}{2} + k_i \zeta]- v^2[\zeta'
\frac{p}{2} + k_i \zeta]^2 -(1-v)\mu^2 -v(\gamma'+\kappa^2)=\nonu=
v(1-v)\left [k^-\left(k^+ +\zeta' {M\over 2} -z_i\zeta{M\over
2}\right) -{M^2 \over 4}z \zeta' - z\zeta{M\over 2}k^-_i -\gamma- 2
\zeta cos \theta \sqrt{\gamma \gamma_i}\right] +\nonu - v^2[\zeta'
\frac{p}{2} + k_i \zeta]^2 -(1-v)\mu^2 -v(\gamma'+\kappa^2)=\nonu=
v(1-v)k^-\left(k^+ +\zeta' {M\over 2} -z_i\zeta{M\over 2}\right)-
\ell_D(v,z,\gamma',\zeta,\zeta',cos\theta) \ee
with (dependence upon
$\gamma_i$, $z_i$ and $\mu$ are understood)
\be
\ell_D(v,z,\gamma',\zeta,\zeta',cos\theta)=v(1-v)\left [ {M^2 \over
4}z \zeta' + z\zeta{M\over 2}k^-_i +\gamma+ 2 \zeta cos \theta
\sqrt{\gamma \gamma_i}\right]  + v^2[\zeta' \frac{p}{2} + k_i
\zeta]^2 +\nonu +(1-v)\mu^2 +v(\gamma'+\kappa^2)=
\nonu
=v(1-v)
\left[ {M^2 \over 4}z \zeta'  +\gamma\right]  + v^2({\zeta'}^{2}
\frac{M^2}{4} + \kappa^2 {\zeta}^2) +\nonu +(1-v)\mu^2
+v(\gamma'+\kappa^2)+v(1-v) \zeta \left[zz_i{M^2\over 4}+  2  cos
\theta \sqrt{\gamma \gamma_i} \right]=
\nonu=
v(1-v)\left[ {M^2 \over 4}z \zeta'  +\gamma+\kappa^2 \right]
 + v^2\left[{\zeta'}^{2}
\frac{M^2}{4} + \kappa^2 (1+\zeta^2)\right] +\nonu +(1-v)\mu^2
+v\gamma'+v(1-v) \zeta \left[zz_i{M^2\over 4}+  2  cos
\theta \sqrt{\gamma \gamma_i} \right]
 \ee Let us recall that $p\cdot k_i=0$ leads to
$ k^+_i+k^-_i=0$ and $\gamma_i=M^2 (1-z^2_i)/4 -m^2$ as mentioned in
the previous Appendix. Then the kernel for obtaining the scattering
state becomes
 \be
V^{(L)}_{s}(\gamma,z;\gamma_i,z_i,\gamma',\zeta,\zeta',cos\theta)= \nonu=~-g^2~p^+{i
\over 2(4\pi)^2}\int_{-\infty}^{\infty}{d k^- \over 2\pi}~ {1 \over
\left[(\frac{p}{2}+k)^2-m^2+i\epsilon\right]} ~ {1\over
\left[(\frac{p}{2}-k)^2-m^2+i\epsilon\right]} ~\times \nonu\int_0^1
\frac{v^2 d v}{[A(p,k,\zeta,\zeta',cos \theta)+i\epsilon]^2}
=~-g^2~p^+{i \over 2(4\pi)^2}~{1 \over ({M\over 2}+k^+)~({M\over
2}-k^+)} \int_0^1 v^2~ d v  \int_{-\infty}^{\infty}{d k^- \over
2\pi}~ ~\times \nonu{1 \over \left[(\frac{p}{2}+ k)^- -
(\frac{p}{2}+ k)^-_{on}+i\epsilon/(M/2+k^+)\right]}~ {1
\over\left[(\frac{p}{2}- k)^- - (\frac{p}{2}-
k)^-_{on}+i\epsilon/(M/2-k^+) \right]}\times \nonu {1
\over\left[v(1-v)k^-\left(k^+ +\zeta' {M\over 2} -z_i\zeta{M\over
2}\right)
-\ell_D(v,\gamma,z,\gamma',\zeta,\zeta',cos\theta)+i\epsilon\right]^2}=
\nonu =~-g^2~{i \over 2\pi^2}~{1 \over M^3 (1-z^2)~} \int_0^1 v^2~d
v  \int_{-\infty}^{\infty}{d k^- \over 2\pi}~ ~\times \nonu{1 \over
\left[(\frac{p}{2}+ k)^- - (\frac{p}{2}+ k)^-_{on}+
i2\epsilon/[M(1-z)\right]}~ {1 \over\left[(\frac{p}{2}- k)^- -
(\frac{p}{2}- k)^-_{on}+i2\epsilon/[M(1+z)]) \right]}\times \nonu {1
\over\left\{v(1-v)(\zeta' -z-z_i\zeta)\left[ k^-
-k^-_D+i2\epsilon/[M v(1-v)(\zeta' -z-z_i\zeta)]\right]\right\}^2}=
\ee where \be (\frac{p}{2}+ k)^-_{on}=p^-_{1on}={2(m^2+\gamma) \over
M (1-z)} \nonu (\frac{p}{2}- k)^-_{on}=p^-_{2on}={2(m^2+\gamma)
\over M (1+z)}\nonu
k^-_D={2~\ell_D(v,\gamma,z,\gamma',\zeta,\zeta',cos\theta)\over M
v(1-v)(\zeta' -z-z_i\zeta)}\label{onshellp}\ee
 One has the following three poles ($1 >z>-1$) \be
k_n=(\frac{p}{2}+ k)^-_{on}-{M\over 2}-i\epsilon \nonu
k_p=-(\frac{p}{2}- k)^-_{on}+{M\over 2}+i\epsilon \nonu k_d= k^-_D-
i{\epsilon \over  (\zeta' -z-z_i\zeta)} \ee
If $\zeta'>z+z_i\zeta$
one can close in the upper plane taking the residue at $k_p$, i.e.
\be\int_{-\infty}^{\infty}{d k^- \over 2\pi}~ {1 \over
\left[(\frac{p}{2}+ k)^- - (\frac{p}{2}+ k)^-_{on}+
i\epsilon\right]}~ {1 \over\left[(\frac{p}{2}- k)^- - (\frac{p}{2}-
k)^-_{on}+i\epsilon) \right]} \times \nonu {1
\over\left[v(1-v)(\zeta' -z-z_i\zeta) (k^-
-k^-_D)+i\epsilon\right]^2}=\nonu =-i{1 \over \left[M- (\frac{p}{2}-
k)^-_{on} - (\frac{p}{2}+ k)^-_{on}+ i\epsilon\right]}~
 \times \nonu
{1 \over\left\{v(1-v)(\zeta' -z-z_i\zeta) [{M\over 2}-(\frac{p}{2}- k)^-_{on} -
k^-_D]+i\epsilon\right\}^2}=\nonu=
-i{M \over \left[M^2- 4{(m^2+\gamma)\over (1-z^2) }
+
i\epsilon\right]}~
{1 \over\left[v(1-v)(\zeta' -z-z_i\zeta) ({M\over 2}-{2 \over M}
{m^2+\gamma\over 1+z} -
k^-_D)+i\epsilon\right]^2}\ee
If $z+z_i\zeta>\zeta'$
one can close in the lower plane taking the residue at $k_n$,
i.e.
\be\int_{-\infty}^{\infty}{d k^- \over 2\pi}~
{1 \over \left[(\frac{p}{2}+ k)^- - (\frac{p}{2}+ k)^-_{on}+
i\epsilon\right]}~
{1 \over\left[(\frac{p}{2}- k)^- - (\frac{p}{2}- k)^-_{on}+i\epsilon)
\right]} \times \nonu
{1 \over\left[v(1-v)(\zeta' -z-z_i\zeta)( k^- -k^-_D-i\epsilon)
\right]^2}=
-i{1 \over \left[M- (\frac{p}{2}+ k)^-_{on} - (\frac{p}{2}- k)^-_{on}+
i\epsilon\right]}~
 \times \nonu
{1 \over\left\{v(1-v)(\zeta' -z-z_i\zeta)[ -{M\over 2}+(\frac{p}{2}+ k)^-_{on} -
k^-_D-i\epsilon]\right\}^2}=
\nonu=
-i{M\over \left[M^2-4{ (m^2+\gamma)\over (1-z^2)}
+
i\epsilon\right]}~
{1 \over\left[v(1-v)(\zeta' -z-z_i\zeta)~({M\over 2}-  {2 \over M} {m^2+\gamma
\over (1-z)} +
k^-_D+i\epsilon)\right]^2}\ee
Then, by exploiting Eq. (\ref{onshellp}) one has
\be
V^{(L)}_{s}(\gamma,z;\gamma_i,z_i,\gamma',\zeta,\zeta',cos\theta)=
~-{g^2 \over 2(2\pi)^2~(1 -z^2)}~{1 \over \left[M^2- 4 {(m^2+\gamma)\over
(1-z^2)}
+
i\epsilon\right]}~\times \nonu
\int_0^1  d v ~v^2{\cal F}(v,\gamma,z;\gamma',\zeta,\zeta',
cos\theta) =\nonu=
~{g^2 \over 2(4\pi)^2}~{1 \over \left[\gamma +(1-z^2)\kappa^2 +z^2 m^2
-i\epsilon\right]}
~
\int_0^1  d v ~v^2{\cal F}(v,\gamma,z;\gamma',\zeta,\zeta',
cos\theta)\ee
where
\be
{\cal F}(v,\gamma,z;\gamma',\zeta,\zeta',
cos\theta)=\nonu =
{\theta (\zeta'-z-z_i\zeta) \over \left[ v (1-v)(\zeta'-z-z_i\zeta)\left({M^2\over 4}-
{m^2+\gamma \over 1+z}\right)- \ell_D(v,\gamma,z,\gamma',\zeta,\zeta',cos\theta)+i\epsilon\right]^2}+ \nonu +
{\theta (z+z_i\zeta-\zeta') \over\left[v (1-v)(z+z_i\zeta-\zeta')\left({M^2\over 4}-
{m^2+\gamma \over 1-z}\right)- \ell_D(v,\gamma,z,\gamma',\zeta,\zeta',cos\theta)
 +i\epsilon\right]^2}=\nonu=
 {(1+z)^2~\theta (\zeta'-z-z_i\zeta) \over \left[{\cal D}(v, z,z_i,\zeta,\zeta')+
  v (1-v) (1+z)\zeta\left({M^2\over 2} z z_i + 2  cos\theta \sqrt{\gamma \gamma_i}\right)-i\epsilon\right]^2}+ \nonu +
{(1-z)^2~\theta (z+z_i\zeta-\zeta') \over\left[{\cal D}(v, -z,-z_i,\zeta,-\zeta')+
  v (1-v)(1-z)\zeta \left({M^2\over 2} z z_i + 2  cos\theta \sqrt{\gamma \gamma_i}\right)
 -i\epsilon\right]^2}
\label{calf}\ee
with
\be
{\cal D}(v, z,z_i,\zeta,\zeta')=v(1-v)(\zeta'-z-z_i\zeta)\left(m^2+\gamma-(1+z){M^2\over 4}
\right)+ \nonu+(1+z) \ell_D(v,\gamma,z,\gamma',\zeta,\zeta',cos\theta)-
v (1-v)(1+z)\zeta \left({M^2\over 2} z z_i + 2  cos\theta \sqrt{\gamma \gamma_i}\right)
=
\nonu=
v(1-v)\left[(\zeta'-z-z_i\zeta)(\kappa^2+\gamma)
+z^2 {M^2 \over 4} (1+\zeta'-\zeta z_i) +(1+z)(\gamma+\kappa^2)\right] +\nonu
+(1+z) \left\{v^2\left[{\zeta'}^{2}
\frac{M^2}{4} + \kappa^2 (1+{\zeta}^2)\right]+(1-v)\mu^2
+v\gamma'\right\}=
\nonu=
v(1-v)\left[(1+\zeta'-z_i\zeta)(\kappa^2+\gamma)
+ {M^2 \over 4} z^2(1+\zeta'-\zeta z_i) \right] +\nonu
+(1+z) \left\{v^2\left[{\zeta'}^{2}
\frac{M^2}{4} + \kappa^2 (1+{\zeta}^2)\right]+(1-v)\mu^2
+v\gamma'\right\}=
\nonu=X(v,z_i,\zeta,\zeta')\left[\gamma +z^2 m^2+\kappa^2(1-z^2)
+\Gamma(v,z,z_i,\zeta,\zeta',\gamma')\right]
\label{cald}\ee
In Eq. (\ref{cald}), the following notation has been used
\be
X(v,z_i,\zeta,\zeta')=v(1-v)(1+\zeta'-z_i\zeta)
\nonu
\Gamma(v,z,z_i,\zeta,\zeta',\gamma')= { (1+z)\over (1+\zeta'-z_i\zeta)}~\times
\nonu\left\{{v\over (1-v)}\left[{\zeta'}^{2}
\frac{M^2}{4} + \kappa^2 (1+{\zeta}^2)+\gamma'\right]+{\mu^2\over v}
+\gamma'\right\}
\ee
Notice that
\be
 {3 \over 4} \geq X(v, z_i,\zeta,\zeta',)\geq 0~~, ~~
 \quad \quad{3 \over 4} \geq X(v, -z_i,\zeta,-\zeta',)\geq 0
\ee
where the positivity follows from   the presence of the theta
functions in Eq. (\ref{calf}).

Then one can rewrite
\be
{\cal F}(v,\gamma,z;\gamma',\zeta,\zeta',cos\theta)=
{(1+z)^2\over X^2(v,z_i,\zeta,\zeta')}~\times \nonu
{\theta (\zeta'-z-z_i\zeta) \over \left[
\gamma +z^2 m^2+\kappa^2(1-z^2)+
\Gamma(v,z,z_i,\zeta,\zeta',\gamma')
+ Z(z,\zeta,\zeta';z_i)
 C -i\epsilon\right]^2}+ \nonu +
{(1-z)^2\over X^2(v,-z_i,\zeta,-\zeta')} ~\times
\nonu
{ \theta (z+z_i\zeta-\zeta') \over
\left[\gamma +z^2 m^2+\kappa^2(1-z^2)+
\Gamma(v,-z,-z_i,\zeta,-\zeta',\gamma')
+ Z(-z,\zeta,-\zeta';-z_i)
 C
 -i\epsilon\right]^2}
\label{calf1}\ee
where $$C={M^2\over 2} z z_i + 2  cos\theta \sqrt{\gamma \gamma_i}$$ and
\be
Z(z,\zeta,\zeta';z_i)=\zeta~{ (1+z)\over (1+\zeta'-z_i\zeta)}
\label{zbig}\ee
Once more,  the presence of the theta
functions in Eq. (\ref{calf}) helps to find the following constrain
\be
|Z( z,\zeta, \zeta',z_i)|\leq |\zeta| ~~, ~~\quad \quad |Z( -z,\zeta, -\zeta',-z_i)|\leq |\zeta|
\label{limitz}\ee
Notice that one can always put $\epsilon \to X(v,z_i,\zeta,\zeta')~\epsilon$,
given the limits  on $X(v,z_i,\zeta,\zeta')$.

The kernel $V_s^{(L)}$ can be evaluated by slightly elaborating the standard
Feynman trick.
Indeed one can use
\be
{1\over B~A^2}= \lim_{\lambda \to 0}~{1 \over \lambda}
\left [{1 \over BA}~-{1\over B(A+ \lambda)}\right]=
\nonu=\lim_{\lambda \to 0}~{1 \over \lambda}
\left\{
\int_0^1 d\xi ~{1\over \left[ B-\xi (B-A) \right]^2}-
\int_0^1 d\xi ~{1\over \left[ B-\xi (B-A) +\xi\lambda\right]^2}\right\}
\label{newtri}\ee
By applying the above relation with the following identifications
\be
A_{\pm}=\gamma +z^2 m^2+\kappa^2(1-z^2)+
\Gamma(v,z,z_i,\zeta,\zeta',\gamma')
+ Z(z,\zeta,\zeta';z_i) C
 -i\epsilon
 \nonu
 B_{\pm}=\gamma +z^2 m^2+\kappa^2(1-z^2)-i\epsilon
\ee
one gets
\be
V^{(L)}_{s}(\gamma,z;\gamma_i,z_i,\gamma',\zeta,\zeta',cos\theta)= ~- ~
~{g^2 \over 2(4\pi)^2}~
 \times \nonu\left [{(1+z)\over (1+\zeta'-z_i\zeta)}
~\theta (\zeta'-z-z_i\zeta)~{\cal H}'(\gamma,z,z_i;\gamma',\zeta,\zeta',cos\theta,\mu^2)
+\right. \nonu \left . +{(1-z)\over (1-\zeta'+z_i\zeta)}
~\theta
(z-\zeta'+z_i\zeta)~{\cal H}'(\gamma,-z,-z_i;\gamma',\zeta,-\zeta',cos\theta,\mu^2)\right]
\label{vls}\ee
where
\be
{\cal H}'(\gamma,z,z_i;\gamma',\zeta,\zeta',cos\theta,\mu^2)=\nonu =
\lim_{\lambda \to 0}~{1 \over \lambda}
\left [ ~{\cal H}(\gamma,z,z_i;\gamma',\zeta,\zeta',cos\theta,\mu^2,\lambda)
-{\cal H}(\gamma,z,z_i;\gamma',\zeta,\zeta',cos\theta,\mu^2,0)\right]
\ee
with
\be{\cal H}(\gamma,z,z_i;\gamma',\zeta,\zeta',cos\theta,\mu^2,\lambda)=
{(1+z)\over (1+\zeta'-z_i\zeta)}
\int_0^1 {d v \over (1-v)^2}\int_0^1 d\xi ~\times
 \nonu {1\over \left\{ \gamma +z^2 m^2+\kappa^2(1-z^2)+\xi\left[
\Gamma(v,z,z_i,\zeta,\zeta',\gamma')
+ Z(z,\zeta,\zeta';z_i)
C +\lambda\right] -i\epsilon\right\}^2}=\nonu
 ={(1+z)\over (1+\zeta'-z_i\zeta)}
\int_0^1 {d v \over (1-v)^2}~\int_0^1 d\xi
\int^\infty_{-\infty}
 d\gamma'' \times \nonu
{\delta \left[\gamma'' - \xi \Gamma(v,z,z_i,\zeta,\zeta',\gamma')
-\xi \lambda\right]\over \left[\gamma +z^2 m^2+
\kappa^2(1-z^2)+\gamma''+\xi
 Z(z,\zeta,\zeta';z_i)C
-i\epsilon   \right]^2} \ee
It should be pointed out that, in order to obtain Eq. (\ref{vls}), the limit on
$\lambda$ has been exchanged with the integral over $v$.

By inserting the following change of variable
\be
z'=\xi Z(z,\zeta,\zeta';z_i)
\ee
such that $|z'|\leq |\zeta|\leq 1$ from Eq. (\ref{limitz}), one gets
\be
{\cal H}(\gamma,z,z_i;\gamma',\zeta,\zeta',cos\theta,\mu^2,\lambda)
={(1+z)\over (1+\zeta'-z_i\zeta)}
\int_0^1 {d v \over (1-v)^2}
\int^\infty_{-\infty}
 d\gamma'' \int_{-1}^1 {dz'\over Z(z,\zeta,\zeta';z_i)} ~\times \nonu
 { \theta(z')~\theta( Z(z,\zeta,\zeta';z_i)-z') - \theta(-z')~
 \theta(z'- Z(z,\zeta,\zeta';z_i)) \over \left[\gamma +z^2 m^2+
\kappa^2(1-z^2)+\gamma''+z'C
-i\epsilon   \right]^2}~
\delta \left[\gamma'' - z'
{\Gamma(v,z,z_i,\zeta,\zeta',\gamma')+ \lambda \over
Z(z,\zeta,\zeta';z_i)}\right]=
\nonu={(1+z)\over (1+\zeta'-z_i\zeta)}
\int_0^1 {d v \over (1-v)^2}
\int^\infty_{-\infty}
 d\gamma'' \int_{-1}^1 dz'{|Z(z,\zeta,\zeta';z_i)|\over Z(z,\zeta,\zeta';z_i)}
~\times \nonu
 \delta \left[\gamma''Z(z,\zeta,\zeta';z_i) - z'
\Gamma(v,z,z_i,\zeta,\zeta',\gamma') - z' \lambda \right]~\times \nonu
 { \theta(z')~\theta( Z(z,\zeta,\zeta';z_i)-z') - \theta(-z')~
 \theta(z'- Z(z,\zeta,\zeta';z_i)) \over \left[\gamma +z^2 m^2+
\kappa^2(1-z^2)+\gamma''+z'C
-i\epsilon   \right]^2}
\ee
Finally, let us perform the change $v \to y/1+y$, then one has
\be
{\cal H}(\gamma,z,z_i;\gamma',\zeta,\zeta',cos\theta,\mu^2,\lambda)=
{(1+z)\over (1+\zeta'-z_i\zeta)}
\int_0^\infty {d y}
\int^\infty_{-\infty}
 d\gamma'' ~\times \nonu\int_{-1}^1 dz' ~\delta
 \left[\gamma''Z(z,\zeta,\zeta';z_i) - z'
\tilde\Gamma(y,z,\zeta,\zeta',\gamma')- z' \lambda \right]\times \nonu
 { \theta(z')~\theta( Z(z,\zeta,\zeta';z_i)-z') + \theta(-z')~
 \theta(z'- Z(z,\zeta,\zeta';z_i)) \over \left[\gamma +z^2 m^2+
\kappa^2(1-z^2)+\gamma''+z'C
-i\epsilon   \right]^2}=\nonu=
\int^\infty_{-\infty}
 d\gamma'' \int_{-1}^1 dz' ~
 { \theta(z')~\theta( Z(z,\zeta,\zeta';z_i)-z') + \theta(-z')~
 \theta(z'- Z(z,\zeta,\zeta';z_i)) \over \left[\gamma +z^2 m^2+
\kappa^2(1-z^2)+\gamma''+z'C
-i\epsilon   \right]^2}~\times \nonu
 \int_0^\infty {d y}~\delta \left[\gamma''~\zeta - z'
\tilde\Gamma(y,\zeta,\zeta',\gamma')- z'{(1+\zeta'-z_i\zeta)\over (1+z)}
 \lambda \right]=
 \nonu=
\int^\infty_{-\infty}
 d\gamma'' \int_{-1}^1 {dz'\over z'} ~
 { \theta(z')~\theta( Z(z,\zeta,\zeta';z_i)-z') - \theta(-z')~
 \theta(z'- Z(z,\zeta,\zeta';z_i)) \over \left[\gamma +z^2 m^2+
\kappa^2(1-z^2)+\gamma''+z'C
-i\epsilon   \right]^2}~\times \nonu
 \int_0^\infty {d y}~\delta \left[{1 \over y}
  \left( y^2~{\cal A}  +y ~{\cal B} +\mu^2\right) \right]
\label{calf2}\ee
where
\be
\tilde\Gamma(y,\zeta,\zeta',\gamma' )=~y \left[{\zeta'}^{2}
\frac{M^2}{4} + \kappa^2 (1+{\zeta}^2)\right]+{(1+y) \over y}\mu^2
+ (1+y)\gamma'
\ee
and the argument of the delta has been  recast in the following form
\be
\tilde\Gamma(y,\zeta,\zeta',\gamma')+ {(1+\zeta'-z_i\zeta)\over (1+z)}
 \lambda -\gamma''~{\zeta\over z'}=\nonu=
 {1 \over y} \left[y^2~{\cal A}(\zeta,\zeta',\gamma',\kappa^2)  +y~
 {\cal B}(z,z',z_i,\zeta,\zeta',\gamma',\gamma'',\mu^2,\lambda) +\mu^2\right]
\ee
with
\be
{\cal A}(\zeta,\zeta',\gamma',\kappa^2)={\zeta'}^{2} \frac{M^2}{4} +
 \kappa^2 (1+{\zeta}^2)+\gamma'
\nonu
{\cal B}(z,z',z_i,\zeta,\zeta',\gamma',\gamma'',\mu^2,\lambda)=\mu^2 +\gamma'-
\gamma'' {\zeta \over z'} +\lambda {(1+\zeta'-z_i\zeta)\over (1+z)}
\ee
The integral over $y$ is readily done, obtaining
\be
{\cal H}(\gamma,z,z_i;\gamma',\zeta,\zeta',cos\theta,\mu^2,\lambda)=
\int^\infty_{-\infty}
 d\gamma'' \int_{-1}^1 {dz' \over z'} ~\times \nonu
 \theta\left( {1+z \over  1 +\zeta' -z_i\zeta}-{z'\over \zeta}\right) { \theta(z')~\theta(\zeta) -
\theta(-z')~\theta(-\zeta) \over \left[\gamma +z^2 m^2+
\kappa^2(1-z^2)+\gamma''+z'C
-i\epsilon   \right]^2}~\times \nonu
\sum_{i=\pm} \theta(y_i(\lambda))~{y^2_i(\lambda) \over
|y^2_i(\lambda){\cal A}(\zeta,\zeta',\gamma',\kappa^2) -\mu^2|} 
\label{calh}\ee
where we have  used Eq. (\ref{zbig}) (exploiting the theta functions for
determining the signs of $\zeta$) and 
\be
\delta(f(y))= \sum_i {\delta(y-y_i) \over |f'(y_i)|}
\ee
with $f(y_i)=0$ and $f'(y_i)$ the derivative evaluated at $y=y_i$. Notice that
$\theta(y_i(\lambda))$ comes from the lower extrema in the integral on $y$.
Moreover, for simplifying the notation,
 we have dropped    the dependence  upon some variables in $y_{\pm}$,
that represent the two solutions of the second order equation
\be
 y^2~{\cal A}(\zeta,\zeta',\gamma',\kappa^2) +y~
 {\cal B}(z,z',z_i,\zeta,\zeta',\gamma',\gamma'',\mu^2,\lambda) +\mu^2=0
\label{eqy}\ee
Once ${\cal H}(\gamma,z,z_i;\gamma',\zeta,\zeta',cos\theta,\mu^2,\lambda)$ is
formally obtained,  its derivative with respect to
$\lambda$, at $\lambda=0$, can be evaluated as follows
\be
{\cal H}'(\gamma,z,z_i;\gamma',\zeta,\zeta',cos\theta,\mu^2)=
\int^\infty_{-\infty}
 d\gamma'' \int_{-1}^1 {dz'\over z'} ~\times \nonu
\theta\left( {1+z \over  1 +\zeta' -z_i\zeta}-{z'\over \zeta}\right) { \theta(z')~\theta(\zeta) -
\theta(-z')~\theta(-\zeta)
 \over  \left[\gamma +z^2 m^2+
\kappa^2(1-z^2)+\gamma''+z'C
-i\epsilon   \right]^2}~\times \nonu
\sum_{i=\pm} {\partial \over \partial \lambda} y_i(0)~
\left\{ \delta(y_i(0)) ~
{y^2_i(0) \over |y^2_i(0){\cal A}(\zeta,\zeta',\gamma',\kappa^2)
 -\mu^2|}+ \right . \nonu \left.
  -\theta(y_i(0)) ~ 2\mu^2 {  y_i(0)
   \over (y^2_i(0){\cal A}(\zeta,\zeta',\gamma',\kappa^2)
 -\mu^2)|y^2_i(0){\cal A}(\zeta,\zeta',\gamma',\kappa^2)
 -\mu^2|}\right\}
 =\nonu=
\int^\infty_{-\infty}
 d\gamma'' \int_{-1}^1  dz'{ {\cal Q}'(z,z_i;\gamma',\gamma'',z',\zeta,\zeta',cos\theta,\mu^2) \over  \left[\gamma +z^2 m^2+
\kappa^2(1-z^2)+\gamma''+z'C
-i\epsilon   \right]^2}
\label{iprime}\ee
where
\be
{\cal Q}'(z,z_i;\gamma',\gamma'',z',\zeta,\zeta',\mu^2)=
\theta\left( {1+z \over  1 +\zeta' -z_i\zeta}-{z'\over \zeta}\right) { \theta(z')~\theta(\zeta) -
\theta(-z')~\theta(-\zeta)\over z'}~\times \nonu
\Lambda\left(z,{z'\over \zeta},\zeta,\zeta';\gamma',\gamma'';z_i,\mu^2\right)
\label{qprimeap}\ee
with
\be\Lambda\left(z,{z'\over \zeta},\zeta,\zeta';\gamma',\gamma'';z_i,\mu^2\right)
= \sum_{i=\pm} {\partial \over \partial \lambda} y_i(0)~\left\{ \delta(y_i(0)) ~
{y^2_i(0) \over |y^2_i(0){\cal A}(\zeta,\zeta',\gamma',\kappa^2)
 -\mu^2|}+ \right . \nonu \left.
  -\theta(y_i(0)) ~ { 2\mu^2~ y_i(0)
  \over (y^2_i(0){\cal A}(\zeta,\zeta',\gamma',\kappa^2)
 -\mu^2)|y^2_i(0){\cal A}(\zeta,\zeta',\gamma',\kappa^2)
 -\mu^2|}\right\}
 \ee
 and
 \be
y_\pm(0)=
{1 \over 2{\cal A}(\zeta,\zeta',\gamma',\kappa^2)} ~\times \nonu
 \left[ -{\cal B}(z,z',z_i,\zeta,\zeta',\gamma',\gamma'',\mu^2,0)
 \pm
\sqrt{{\cal B}^2(z,z',z_i,\zeta,\zeta',\gamma',\gamma'',\mu^2,0)
-  4\mu^2~ {\cal
A}(\zeta,\zeta',\gamma',\kappa^2)} \right]
\nonu
{\partial \over \partial \lambda} y_i(0)=~\mp~
{(1+\zeta'-z_i\zeta)\over (1+z)}~~ 
~{ y_\pm(0)\over
\sqrt{{\cal B}^2(z,z',z_i,\zeta,\zeta',\gamma',\gamma'',\mu^2,0)
- 4 \mu^2 {\cal A}(\zeta,\zeta',\gamma',\kappa^2)}} \ee
Let us recall that $y_i(0)$ still depends upon $z,z_i,\zeta$ and $\zeta'$, and
therefore the proper changes, dictated by the arguments of ${\cal Q}'$,  
must be performed in order to get the final
expression for $V_s^{(L)}$.  The necessity of real  solutions for getting non
vanishing Nakanishi amplitude imposes  a constraint on the discriminant of Eq.
(\ref{eqy}), viz 
\be
{\cal B}^2(z,z',z_i,\zeta,\zeta',\gamma',\gamma'',\mu^2,0)
-  4\mu^2~ {\cal
A}(\zeta,\zeta',\gamma',\kappa^2)=\nonu=\left(\mu^2 +\gamma'-
\gamma'' {\zeta \over z'}\right)^2 - 4\mu^2\left[
{\zeta'}^{2} \frac{M^2}{4} +
 \kappa^2 (1+{\zeta}^2)+\gamma'\right]\geq ~0\ee
More constraints can be obtained from the positivity of the solutions $y_i(0)$.
 In particular one has
 \begin{itemize}
 \item
 if ${\cal A}(\zeta,\zeta',\gamma',\kappa^2) \geq 0$,  then one has to
 consider
 ${\cal B}(z,z',z_i,\zeta,\zeta',\gamma',\gamma'',\mu^2,0)\leq -2\mu \sqrt{{\cal A}(\zeta,\zeta',\gamma',\kappa^2)}$.
 In this case, two solution for $y_i(0)$ are allowed;
 \item if ${\cal A}(\zeta,\zeta',\gamma',\kappa^2) <0$,  then 
 ${\cal B}(z,z',z_i,\zeta,\zeta',\gamma',\gamma'',\mu^2,0)\geq 0$.
 Only one solution is allowed.\end{itemize}
 
In conclusion, from Eq. (\ref{vls}) and
using Eq. (\ref{iprime}),  one has
\be
V^{(L)}_{s}(\gamma,z;\gamma_i,z_i,\gamma'',\zeta,\zeta',cos\theta)= ~- ~
~{g^2 \over 2(4\pi)^2}~\int^\infty_{-\infty}
 d\gamma' \int_{-1}^1 {dz'}~\times \nonu{1 \over  \left[\gamma +z^2 m^2+
\kappa^2(1-z^2)+\gamma'+z'\left({M^2\over 2} z z_i + 2  cos\theta \sqrt{\gamma
\gamma_i}\right)
-i\epsilon   \right]^2}
 \times \nonu \left [{(1+z)\over (1+\zeta'-z_i\zeta)}
~\theta (\zeta'-z-z_i\zeta)~
{\cal Q}'(z,z_i;\gamma'',\gamma',z',\zeta,\zeta',\mu^2)
+\right. \nonu \left . +{(1-z)\over (1-\zeta'+z_i\zeta)}
~\theta
(z-\zeta'+z_i\zeta)~
{\cal Q}'(-z,-z_i;\gamma'',\gamma',z'\zeta,-\zeta',\mu^2)\right]
\label{vls1}
\ee
where for the sake of notation adopted in Sect. \ref{seclad} the change $\gamma''
\to \gamma'$ has been performed.


\end{document}